\documentclass[preprint]{aastex}

\shorttitle{Visualizing Astronomical Data with Blender}
\shortauthors{Kent}

\begin{document}

\title{Visualizing Astronomical Data with Blender}

\author{Brian R. Kent}
\affil{National Radio Astronomy Observatory\altaffilmark{1}\\
520 Edgemont Road, Charlottesville, VA, 22903, USA \\
Email: bkent@nrao.edu\\
Web: http://www.cv.nrao.edu/$\sim$bkent/computing/kentPASP.html\\
-\\
Published in the Publications of the Astronomical Society of the Pacific}

\altaffiltext{1}{The National Radio Astronomy Observatory is a facility of the  
National Science Foundation operated under cooperative agreement by  
Associated Universities, Inc. }

\begin{abstract}
Astronomical data take on a multitude of forms -- catalogs, data cubes, images,
and simulations.  The availability of software for rendering high-quality
three-dimensional graphics lends itself to the paradigm of exploring the incredible
parameter space afforded by the astronomical sciences.  The software program
Blender gives astronomers a useful tool for displaying data in a manner used
by three-dimensional (3D) graphics specialists and animators.  The interface to this popular
software package is introduced with attention to features of interest in astronomy.
An overview of the steps for generating models, textures, animations, camera work, and renders is outlined.  An introduction is presented on the methodology for producing animations
and graphics with a variety of astronomical data.  Examples from 
sub-fields of astronomy with different kinds of data are shown with resources
provided to members of the astronomical community.
An example video showcasing the outlined principles and features is provided
along with scripts and files for sample visualizations.

\end{abstract}


\section{Introduction}
\label{intro}

The combination of astronomy and computational sciences plays an important role in how members of the astronomical community visualize their data.  Larger detectors in the optical/infrared (IR), increasing bandwidth in radio/millimeter interferometers, and large \textit{N} astrophysical simulations drive the need for visualization solutions.  These visualization purposes include exploring the dynamical phase space of data cubes, surface mapping, large catalogs, and volumetric rendering.  The use of 3D computer graphics in the movie, television, and gaming industry has led the development of useful computer algorithms for optimizing the display of complex data structures while taking advantage of new hardware paradigms like utilizing graphic processing units (GPUs).  Melding the exciting scientific results with state of the art computer graphics not only helps with scientific analysis and phase space discovery, but also with graphics for education and public outreach. Exploratory data visualization can offer a
complementary tool next to traditional statistics \citep{2012AN....333..505G}.  The work presented here is motivated by a dearth of clear instruction for astronomers in 3D graphics, animation, and accessible visualization packages.  The visual impact of astronomy cannot be understated -- for both the community of scientists and how to present our results to the public with higher accuracy.

We present a practical overview for astronomers to the capabilities of the software program Blender -- an open-source 3D graphics and animation package.  We take the approach of describing the features that will be of interest to various parties in astronomy -- for both research visualization and presentation graphics.  We aim to describe methods of data modeling, texturing, lighting, rendering, and compositing.  Section~\ref{vizdata} discusses some of the relevant history of data visualization in the sciences.  Section~\ref{blender} introduces the software, interface, and workflow.  Section~\ref{workflowsession} describes the methodology of a sample workflow session using an astronomical position-position-frequency data cube at radio frequencies.  Section~\ref{examples} describes examples from different fields within astronomy.  Section~\ref{summary} summarizes the method overview and includes a demonstration
video outlining the foundations and examples built upon in this paper.

\section{Visualizing Data in the Sciences}
\label{vizdata}

Three-dimensional visualization allows for the exploration of multiple dimensions of data and seeing aspects of phase space
that may not be apparent in traditional two-dimensional (2D) plotting typically used in analysis.  Visualizing scientific data
helps to complement the narrative of an observation or experiment.  A number of studies have reviewed the progression and evolution of the field in software engineering and the sciences \citep{journals/ac/StaplesB99, doi:10.1146/annurev.psych.52.1.305, Diehl:2007:SVV:1209814, Teyseyre:2009:OSV:1477065.1477374}.  Beginning with the first graphs and charts of analog mapping data \citep{nla.cat-vn108090}, data visualization has grown as technology now allows us to explore multiple dimensions of 
parameter spaces.  \citet{Friendly:06:hbook} gives an overview of visual thinking and data presentation in mathematics, statistics, and the sciences.
\citet{Hansen2005} describe the computer graphics nomenclature, techniques of volume rendering, frameworks, and issues of perception in visualization.  

The field of visualization in the physical sciences, including astronomy, faces challenges of scalability \citep{4015437}. 
Technology only brings us part of the way toward successfully understanding how to best visualize scientific data -- understanding the theory of light, reflection, and color in graphics is needed to produce useful results \citep{Frenkel:1988:ASV:42372.42373}.  In addition, using a workflow and framework that maximizes the use of layers and node composition can optimize the visualization of data \citep{Birn:2000:DLR:556898}.  The goals of scientific visualization are twofold, as are the underlying challenges -- we want
to analyze quantitatively with high accuracy and precision our experiments, but at the same time produce stunning visuals that convey results beyond the scientific community \citep{Munzner2006}.

Notable usage of scientific visualization, particularly with Blender, include the animation of maps from geographic information systems\citep[GIS;][]{scianna2013}, biology \citep{Autin:2012:UUC:2412364.2412538}, and protein models \citep{2010arXiv1009.4801Z}.  Algorithm development in medical imaging has paved the way 
in visualizing tomography and magnetic resonance \citep{Lorensen:1987:MCH:37402.37422}. \citet{Drebin:1988:VR:378456.378484}, \citet{Elvins:1992:SAV:142413.142427}, and \citet{Yagel96classificationand} list early reviews, developments, and taxonomy of algorithms for volume rendering and viewing, many of which started in the film and animation industry.  While applications may be particular to specific science sub domains, the data and visualization
goals are often similar and well aligned \citep{2007ASPC..376..621B}.  \citet{CGF:CGF3184} reviews the literature across the different physical sciences, breaking down visualization into the areas
of 2D, 3D, and time-series data analyses.

With the advent of large sky surveys across the electromagnetic spectrum, short cadence 
temporal observations, and high-resolution simulations, astronomy now requires innovative use 
of software to analyze data \citep{2011ApJS..192....9T}.  
These data and methodologies can be catalogs \citep{Buddelmeijer,  931430, 2011AAS...21840822T}, images \citep{1996A&AS..117..393B, 2004PASP..116..133L, 2011ASPC..442..169L}, multidimensional datacubes \citep{2008ASPC..394..339D, 2011ASPC..442..625K}, spectra and time-series \citep{2005ASPC..347..143L, 2006ASPC..359..285B, 2008CQGra..25r4025M}, and simulations \citep{2001ASPC..238..499T}.   New techniques are utilizing high-performance computing and access to distributed data \citep{Strehl02relationship-basedclustering, 2007PASP..119..898C}.  Volume rendering, lighting, and shading play a key roll in how data are presented \citep{485155, Direct_Interval_Volume_Visualization, tvcg11_dvr}.  Still, other experiments require new algorithms or innovative uses
of existing hardware to push through challenging roadblocks that arise with new scientific paradigms \citep{nebulae-vis2012,3c40fd5dd5}. \citet{2011PASA...28..150H} review different approaches
in astronomy looking to the future of the field as data rates increase.

This work examines how astronomers can use the 3D graphics software Blender to visualize different types of astronomical data for purposes of both research and education and public outreach.
We briefly compare Blender to other 3D packages in section \ref{comparisonsoftware}.
Using a 3D graphics package gives control over the 4th dimension of many datasets -- namely
time when animating videos.  In addition, cinematic camera mechanisms and object manipulation used during rendering can give control other packages cannot \citep{Balakrishnan:1999:EBC:302979.302991}.
While there is a wealth of specialized software libraries and algorithms available,
we choose to introduce Blender to astronomers for its versatility and flexibility in handling
different types of visualization scenarios in astronomy.

\section{Blender}
\label{blender}

Blender is a software package aimed at supporting high-resolution production quality 3D graphics, modeling, and animation.  The open-source software is maintained by the Blender Foundation with a large user base and over 3.4 million downloads per year\footnote{Available from http://www.blender.org/}.  The official release under the GNU GPL is available for Linux (source and pre-built binaries), Mac OS X, and Windows on 32-bit and 64-bit platforms.   The graphical user interface (GUI) is designed and streamlined around an animator's production workflow with a Python application program interface (API) for scripting.  The API is advantageous in astronomy for loading data through PYFITS or CFITSIO as well as the growing collection of Python modules available for astronomy.  The final composited output can be high-resolution still frames or video.  While the package has numerous uses for game theory and design, video editing, and graphics animation, here we focus on the components most applicable to scientific visualization in astronomy.  As an existing stand alone software package Blender provides a framework for both astronomical data exploration as well as education and public outreach.  Motivations for usage in astronomy can be for exploratory data analysis by rendering
volumetric data cubes, projecting and mapping onto mesh surfaces, 3D catalog exploration,
and animating simulation results.  Blender meets the requirements for both audiences -- that
of rigorous scientific accuracy in addition to the fluid cinema animations that bring
the visual excitement of astronomy to the public.

\subsection{Interface and Workflow}

The main Blender interface is shown in Figure~\ref{maininterface} and shows the object toolbar, main 3D viewport, transformation toolbar, data and model outliner, properties panel, and animation time line.
A typical session workflow in 3D animation consists of the following steps:\\

\begin{figure}
\epsscale{1.0}
\plotone{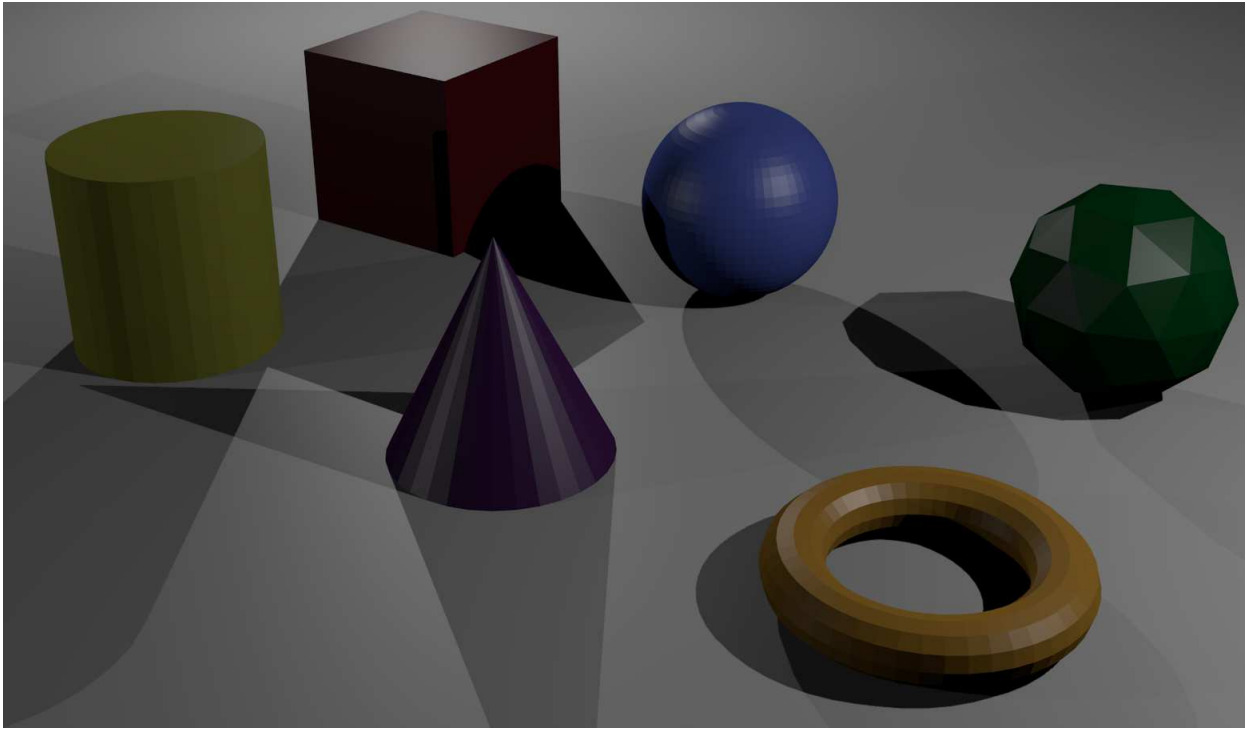}
\caption{Six basic mesh primitives.  This example scene shows faceted versions of a cube, UV-sphere, icosahedron, cylinder, cone, and torus, each colored with a different material.  These are simple objects upon which more complex models can be built.\label{meshexamples}}
\end{figure}

\begin{figure}
\epsscale{1.0}
\plotone{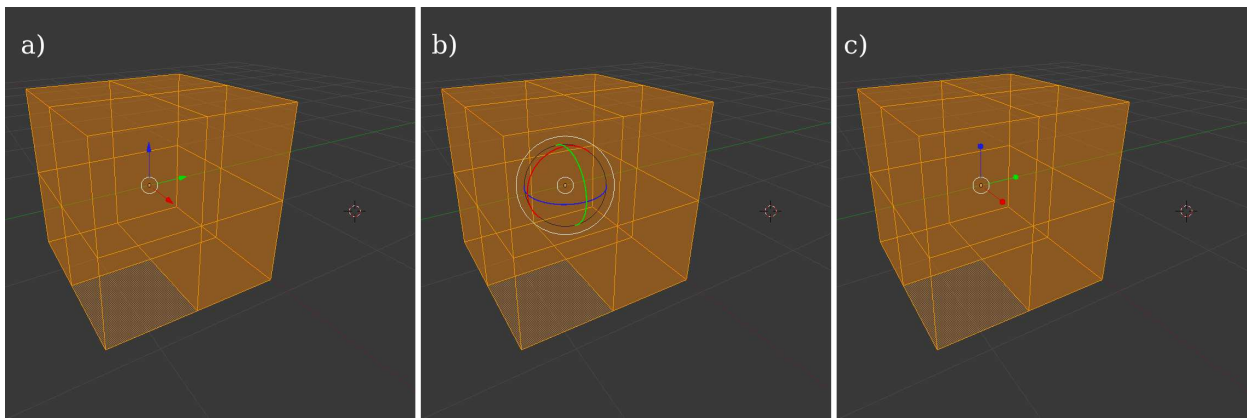}
\caption{Three different control handlers of the interface. \textit{Red}, \textit{blue}, and \textit{green} controls correspond to \textit{x}, \textit{y}, and \textit{z}.  (\textit{a}) Translation widget arrows, (\textit{b}) Meridians of the rotation widget. (\textit{c}) Box handlers of the scaling widget.  \label{controls}}
\end{figure}

\noindent \textbf{Modeling.}  The user generates models, called meshes, that conform to the type of data they wish to view (Figure~\ref{meshexamples}).  Fundamental mesh primitives can act as data containers (for 3D data cubes), volumes (for gas flow simulations), 3D data points (for \textit{N}-body simulations) or surfaces (for planetary surface maps).  These shapes can morph and change their location, rotation, and scale during animation (Figure~\ref{controls}).  Each mesh, no matter the size, consists of three basic components -- vertices, edges, and faces.  As the properties of these basic components are modified the data visualization begins to take shape.

\begin{figure}
\epsscale{1.0}
\plotone{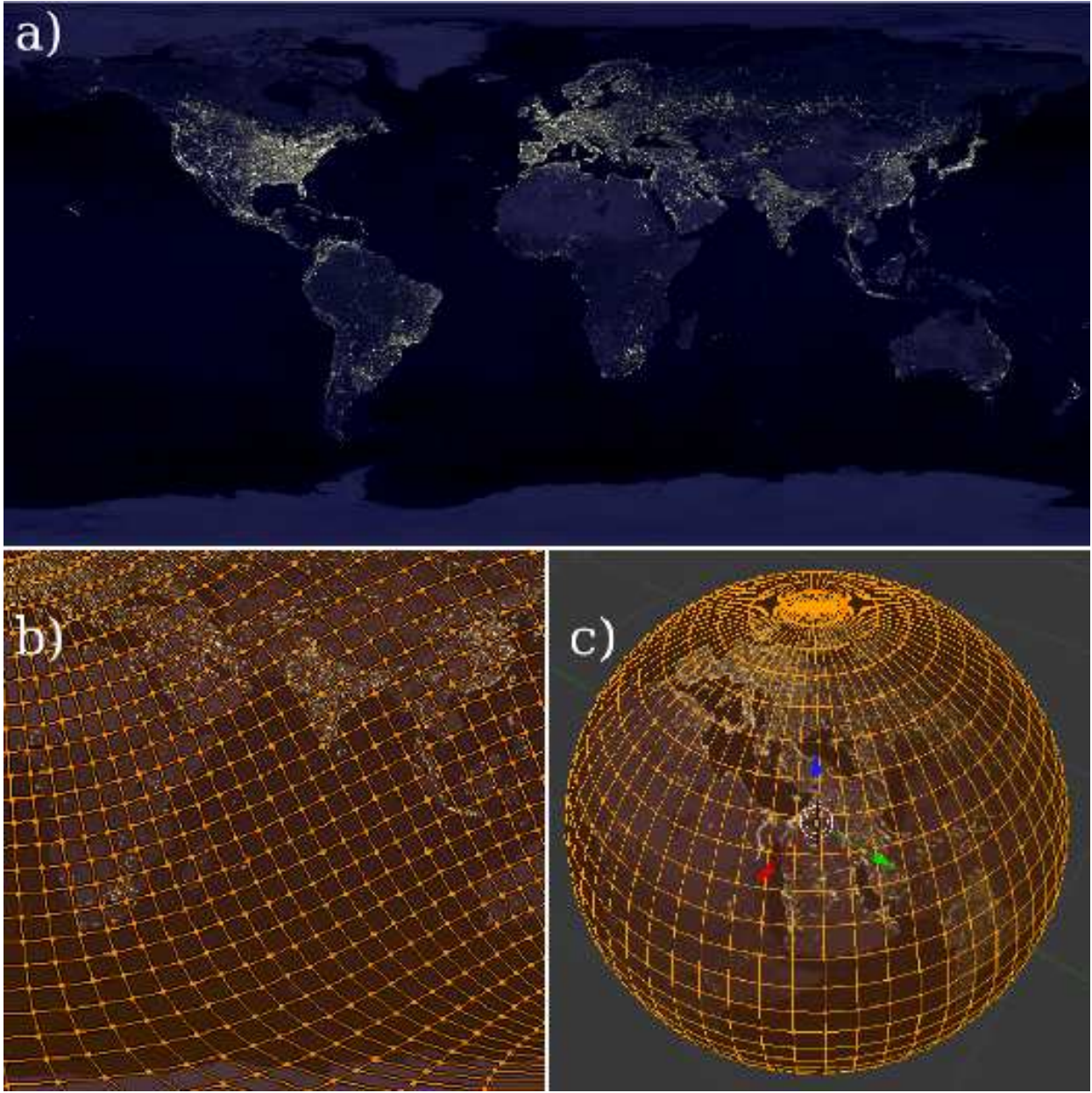}
\caption{Example of UV mapping with data from the Defense Meteorological Satellite Program (DMSP) Operational Linescan System \citep[OLS;][]{Elvidge1997}, shown in (\textit{a}).  The surface of the UV-sphere is mapped onto the cylindrical projection of the nighttime view of the Earth in (\textit{b}); (\textit{c}) shows the final result of the mapping. \label{uvmapping}}
\end{figure}

\noindent \textbf{Texturing and Mapping.}  Texturing is no longer limited to simple repetitive patterns of bit-mapped images on surfaces.  Such surface texture mapping can be useful for planetary surfaces.  3D animation packages use a technique called UV-unwrapping (not to be confused with \textit{u-v} coordinates used in interferometry).  This technique projects surfaces of 3D models onto images (Figure~\ref{uvmapping})\footnote{Images available at http://earthobservatory.nasa.gov}.  Bump mapping is another useful utility allows the simulation of surface differences using perturbations of the surfaces' normals during the render stage, where a full 3D model would have been needed otherwise \citep{Blinn:1978:SWS:965139.507101}.  Blender does not possess the native ability to process astronomical coordinate headers specified in FITS file headers.  Further processing through a Python script would be required in order to reorient surface models
to the correct map projection; however, the capabilities exist for such endeavors, made easier by the Python API and supporting astronomical libraries\footnote{http://stsdas.stsci.edu/astrolib/pywcs/}.

Astronomy data that is multidimensional has signal buried in noise and must use volumetric texturing.  Human perception has the uncanny ability to act as an amazing signal extractor in one-dimensional spectra or two-dimensional images.  However, seeing through a data cube requires us to allow signal emission to pass through a specified noise level.  The computation of a ray as it passes in and out of one volumetric pixel (a voxel) to the next is given by a simple transfer function \citep{Levoy:1988:DSV:44650.44652}:
\begin{equation}
C_{out, RGB}(u_i , v_j) = C_{in, RGB}(u_i , v_j)(1 - \alpha (x_i,y_j,z_k)) + c (x_i,y_j,z_k) \alpha (x_i,y_j,z_k)
\end{equation}

where $C_{out, RGB}$~and~$C_{in, RGB}$ are the red, green and blue (RGB) values
of entry and exit at pixel $(u_i , v_j)$, and $\alpha$ is the opacity.  $c (x_i,y_j,z_k)$ represents the color values at the point of interaction.  Viewing this signal while "seeing through" the noise requires a combination of opacity channels (called alpha in the graphics industry) and ray tracing.  The output pixel projected onto the camera plane is a composited red, green blue, and alpha (RGBA) rendered image \citep{Porter:1984:CDI:964965.808606}.  In addition astrophysical visualization, this paradigm exists in other fields including medical imaging and examining large microscopy image stacks \citep{2011ApJS..197...18F, 568151, Peng2010}.
The paradigm is functionally the same as frequency channels in an astronomical data cube obtained with a radio telescope.
Noncube voxel shapes present an interesting challenge for rendering algorithms, as some 3D models (or FITS data cube projections in astronomy) are not fundamentally a simple \textit{x-y-z} orthogonal projection \citep{Reniers2008}.  The current \textit{VoxelData}\footnote{\scriptsize{$\rm{blender.org/documentation/blender\_python\_api\_2\_65\_release/bpy.types.VoxelData.html}$}} base class in Blender allows for any mesh shape, including non-cubic, to be containers for voxel textures.  This has potential for more complex map projections in astronomical surveys, large scale structure simulations, adaptive mesh refinement codes (AMR) or tessellation algorithms; it is typically used in animation for web/sponge-like constructs where coordinates can be mapped more easily onto non-cubic surfaces \citep{Crassin2009, Strand2005}.

\noindent \textbf{Lighting.}  This step specifies how light will reflect, refract, and scatter off of physical surfaces.  For astronomical data we will be concerned with how the data are presented on a graphics display that the user will see through the camera viewport.  Various lighting algorithms are available depending on the presentation required \citep{Phong:1975:ICG}.

\begin{figure}
\epsscale{1.0}
\plotone{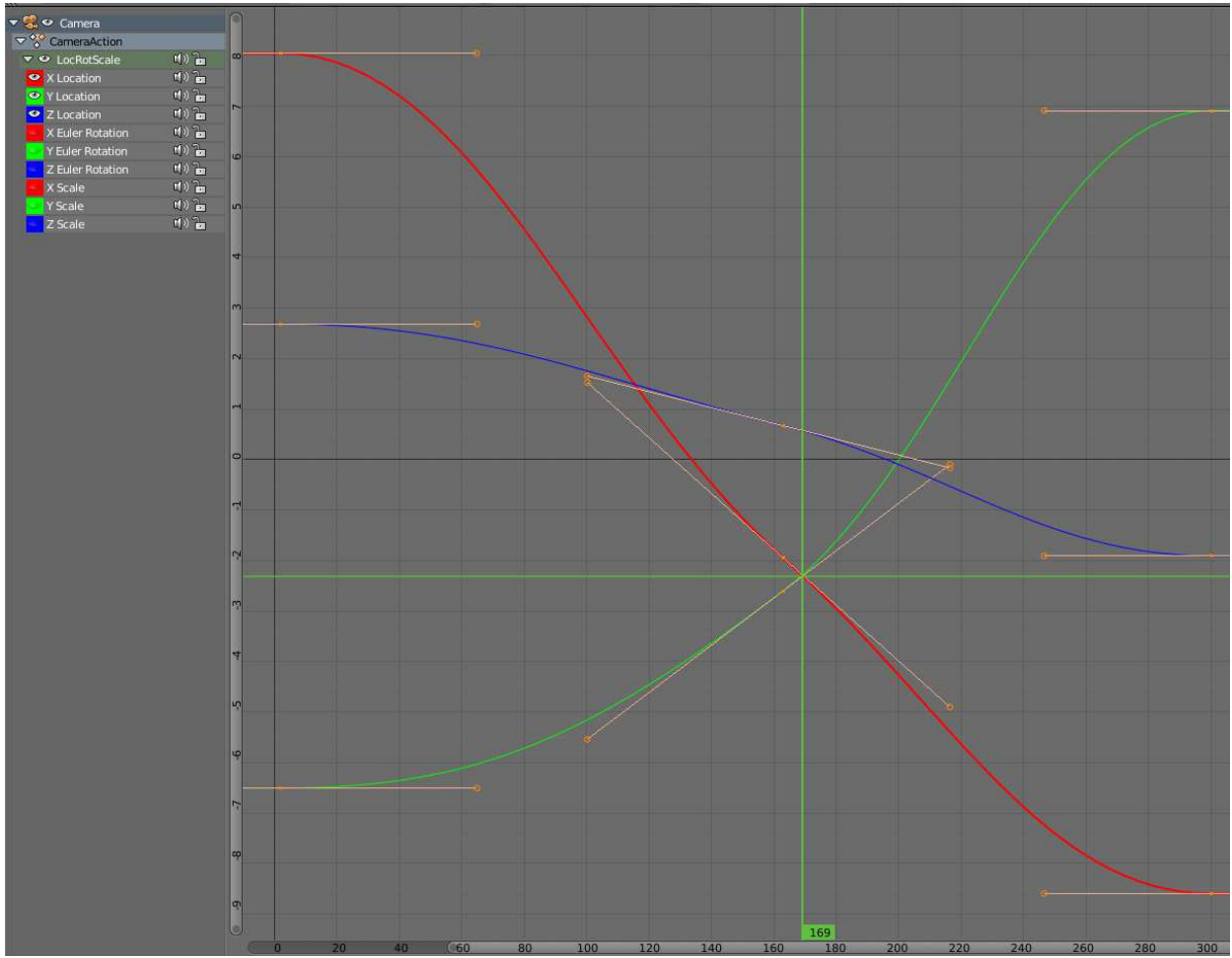}
\caption{Example graph editor session that allows the user to manipulate complex movement curves.  A camera with \textit{red}, \textit{blue} and \textit{green} curves for its \textit{x}, \textit{y}, and \textit{z} location is shown as it tracks an object in a circular orbit.  The representation shown here moved the camera closer to the object at periapsis and then passes through the orbital plane.  The \textit{thin orange lines} tangent to the curves act as control handles for manipulation.  The horizontal axis shows the frame number and the vertical axis shows
the position along the respective \textit{x},\textit{y}, or \textit{z}-axis.  \label{grapheditor}}
\end{figure}

\noindent \textbf{Animation.}  Animation is accomplished by recording a change of state in the mesh models and their associated properties.  This task uses "keyframing" to record location in 3D space, translation, rotation, and relative scaling between frames.  Keyframe pivot points can rapidly increase in number - fortunately there is a built in feature, the "graph editor" that gives a schematic view of movement during animation.  Figure~\ref{grapheditor}~
shows the graphical view of a camera location as it tracks an object during an animation.

\noindent \textbf{Camera Control.}  In addition to animating the meshes, models, and properties, camera movement and rotation, focal length, depth of field can be controlled and keyframed.  This is one of the most useful features of Blender -- giving the user complete control over how to view the generated animation.  The camera can follow a path past a specified object while tracking a given feature.  The camera can fly an orbit around a data cube, and then move in closer to chosen features.  This utility is made easier in the GUI rather than the scripting mode because the user can see where the camera is moving relative to the models.

\noindent \textbf{Rendering.}  Rendering can be handled by a number of internal rendering engines -- the code that takes all the animation movements, textures, and lighting emission and absorption parameters, and generates the frames of animation, as viewed by the camera (or multiple cameras, for multiple rendering passes).  Blender has two engines that are included by default -- called Render and Cycles.  Both serve useful purposes depending on the kind of data visualization being performed.  We will discuss these and other third-party rendering engines in section \ref{renderingengines}.

\noindent \textbf{Compositing.}  Complex data visualization usually benefits from having the aforementioned data models, animation, cameras, and render output separated for ease of manipulation.  A single image in a raster graphics program such as Adobe Photoshop or GIMP is normally separated into multiple layers, channels, and paths to facilitate manipulating transparency, brightness, and saturation.  In the same vein with 3D graphics and animation packages, multiple layers can be combined into the final animation via the "node compositing editor."  Using nodes gives the user a tree-based graph that links layers, models, and any final graphical corrections to the final view output into a map showing the progression on the final composite \citep{duff1985, Brinkmann2008}.  In a fashion similar to layers in astronomical browsing utility Aladin \citep{2011ASPC..442..683B}, separate rendered frames from different source models can be combined.  Final frames can be added together using the "video sequence editor" in a variety of standard graphics formats (PNG, GIF, JPEG, and TIF for single
frames, and AVI, H.264, Quicktime, and MPEG for video among many others).

\begin{figure}
\epsscale{1.0}
\plotone{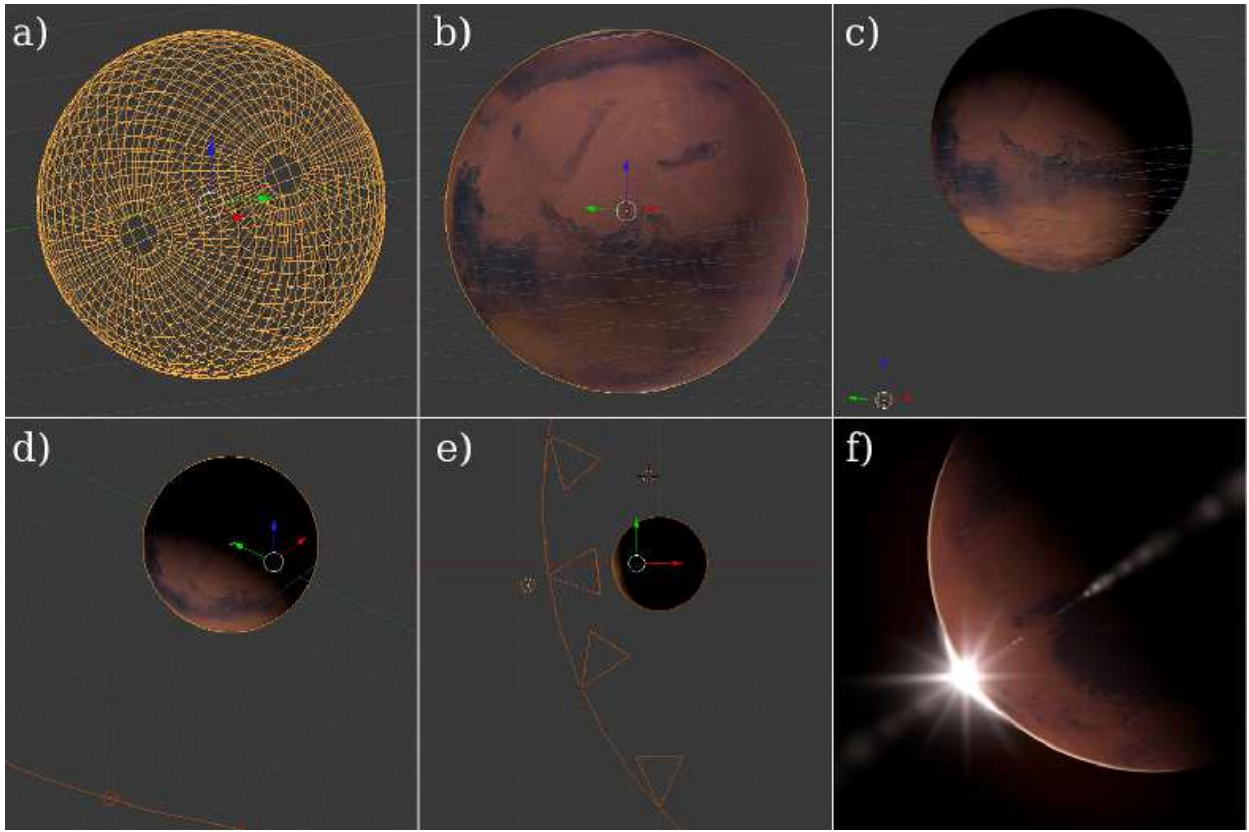}
\caption{Blender workflow process.  (\textit{a}) UV-sphere mesh.  (\textit{b}) Model with Mars data \textit{uv}-mapped onto the mesh.  (\textit{c}) Lighting element being applied.  (\textit{d}) Small curve with the lighting element being animated.  (\textit{e}) Camera animation curve (\textit{lower left in the panel}) with a \textbf{Track To} constraint applied in the \textbf{Properties} panel.  This allows the camera to move along the path but to always be focused on the mesh sphere.  (\textit{f}) Frame of the final render and composite.  \label{workflow}}
\end{figure}

The GUI can change depending on the task at hand (modeling, animation, compositing, or sequencing), and can be fully customized by the user.  Intricate details of using the GUI are beyond the scope of this introductory paper.  The Blender documentation\footnote{Manual available at http://wiki.blender.org/index.php/Doc:2.6/Manual} and workbooks contain numerous tutorials for learning how to navigate the interface \citep{Hess:2010:BFE:1893021}.
Figure~\ref{workflow} shows an outline of the workflow process from modeling through the final render of a planetary surface map using data from the NASA \textit{Viking} Orbiter\footnote{Maps available at http://www.mars.asu.edu/data/mdim\_color/}.

\subsection{Blender Data Structures and Outliner}
\label{blenderdatastructures}

\begin{figure}
\epsscale{1.0}
\plotone{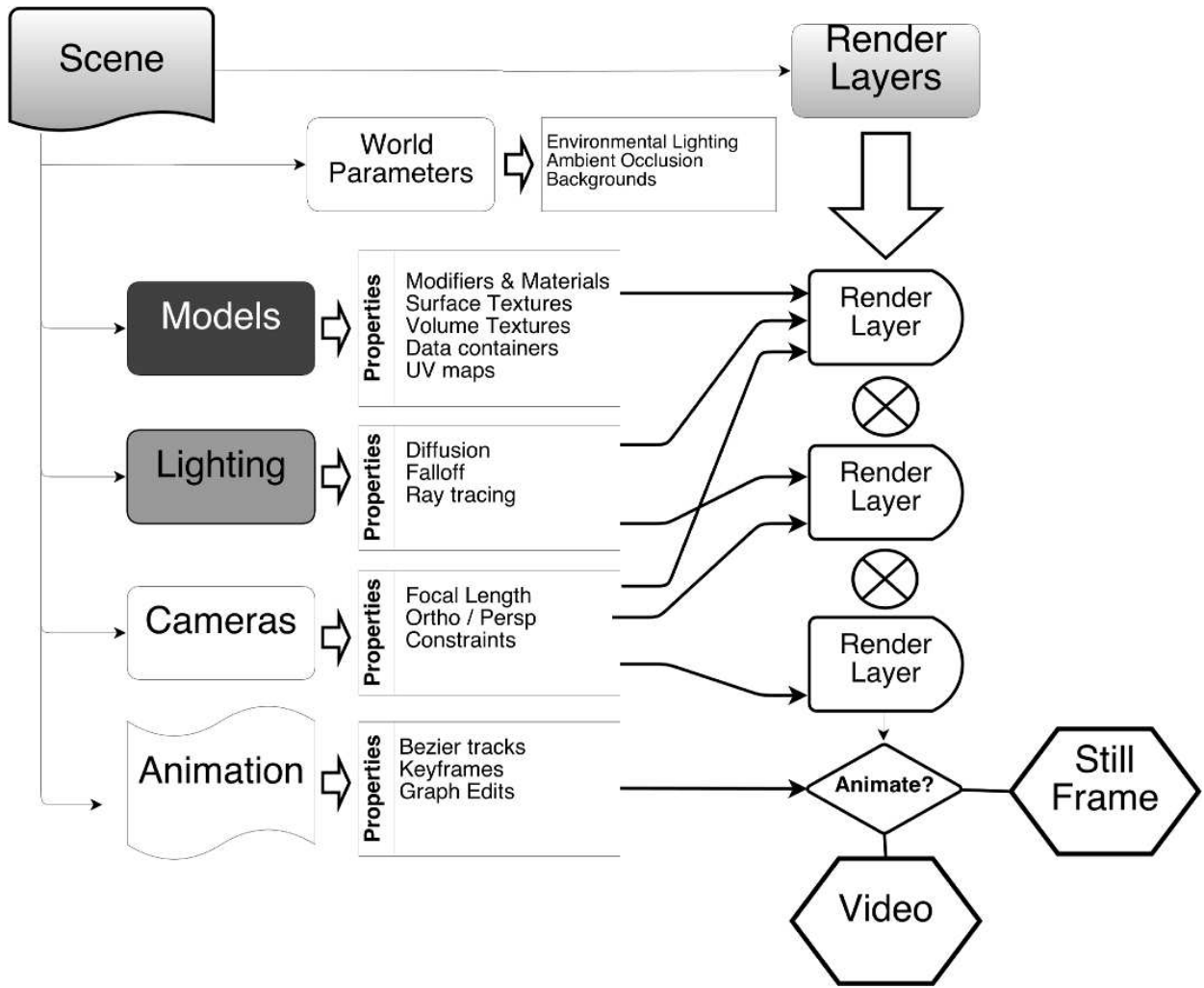}
\caption{Data object structures and flow in Blender.  Model, lighting, lamera, and animation properties are fed into different render layers.   These render layers are then composited into a final video or still frame, with compositing indicated by $\bigotimes$.  In this example, each render layer is captured by cameras, but models only contribute to the first layer shown on the top.  Other layers might include backgrounds, lighting, blurs, or lens flares that are composited after the models are rendered and textured.  \label{flowchart}}
\end{figure}

The Blender core is written in C, with C++ for windowing and Python for the API and scripting functionality.  Native Blender files (*.blend) contain headers with identifiers, pointer sizes, endianness, and versioning.  These headers are followed by data blocks with sub headers containing identifier codes, data sizes, memory address pointers, and data structure indices.  Each block contains data arrays for object names and types (mesh, light, camera, curve, etc.), data types, data lengths, and structures, which may contain sub arrays of the fields.  The data block for an animated scene consists of structures detailing objects and their properties - locations of vertices, edges, faces, and textures, and how those might be affected by a given keyframe.  A sample of how these data structures fit together in the workflow paradigm is shown in Figure~ \ref{flowchart}.  Hovering over any element, button, or menu in the Blender interface gives a context sensitive reference to classes and methods in the API\footnote{API documentation available at http://www.blender.org/education-help/}.  These can be used to write procedural Python scripts for automating tasks within Blender.

\subsection{Rendering Engines}
\label{renderingengines}

Blender can show real time previews of the animated scene in wireframe, solid, or textured mode, allowing the user to plan and anticipate when and where camera placements and movements need to take place.  When the render and composite of all layers is required, the Blender data model is passed to a rendering engine, software that computes the emission, absorption, scattering, reflection, shading, and transport of simulated light onto the projected camera plane.  The following rendering engines can be used with Blender; this list is not exhaustive, but gives the reader a number of open-source references to begin exploring different options.

\noindent \textbf{Render.}  The first generation internal rendering engine is useful for most general purpose data visualizations that astronomers will encounter.  It handles volumetric rendering well and supports GPU processing
(\textit{http://wiki.blender.org/index.php/Doc:2.6/Manual/Render}).

\noindent \textbf{Cycles.}  This second generation rendering engine adds more realistic caustics to the rendering algorithm.  In addition, the rendering path can be configured within the node-based compositor, allowing for rapid prototyping of a scene
(\textit{http://wiki.blender.org/index.php/Doc:2.6/Manual/Render/Cycles}).

\noindent \textbf{LuxRender.}  This open-source third party rendering engine that achieves fast raytracing with GPU processing and uses a progressive rendering algorithm
    (see references in \citet{6183117}; \textit{http://www.luxrender.net/}).

\noindent \textbf{YafaRay.}  This open-source third party raytracing and rendering engine that has the ability to save files to high dynamic range images formats \citep[EXR;][\textit{http://www.yafaray.org/}]{Debevec:1998:RSO:280814.280864}.

In addition, usage of the utility FFmpeg\footnote{Available at http://www.ffmpeg.org} can be used to encode video to popular video formats, frame rates, from production quality high-definition (HD) video to sizes and data compressions more suitable to mobile devices.

\subsection{Data Formats}

Combined with its Python API, Blender is able to import and export a number of data formats useful in astronomy.  In addition to its native "blend" format described in Section \ref{blenderdatastructures}, the formats listed in Table \ref{blenderformats} can be utilized for data input and output.
ASCII tables with fields separated by delimiters can be imported via a number of native Python modules.
FITS images and tables can be imported via PYFITS or CFITSIO onto surface or volumetric texture containers.
JPEG, TIFF, PNG, and GIF images can be read natively by Blender and mapped onto surface textures.
Mesh models can be imported or exported with industry standard Wavefront files,
which consists of ASCII text defining vertex intersections \citep[.OBJ;][]{0083492}.

Single frames can be exported in a variety of image formats including JPEG, TIFF, PNG, GIF, and EXR,
each with full control over binning and compression of the image output \citep{Wallace:1991:JSP:103085.103089}.  
Video output is dependent on the operating system; Windows, Linux, and Mac OS can all export MPEG and AVI video files.  For Quicktime MOV and H.264/MPEG-4 (useful in mobile devices), the aforementioned FFmpeg utility can be used with Linux.
Each format can also be imported into the video sequencer -- a powerful feature of the software
but beyond the scope of this paper.

\subsection{High performance Computing}

Visualizing and animating output for inspection or study can be run as parallel processes or a graphics processing unit \citep[GPU;][]{2012PASA...29..340H}.  Once keyframes have been inserted, frames generated from the render process become independent.  These frames can be rendered through multiple threads on a GPU, multiple cores on a single processor, or on completely independent workstations.  NVidia CUDA and OpenCL are both supported with appropriate hardware \citep{4626815, 5457293}.  Completed animation sequences can be sent to "render farms" that will render and composite the final images or videos, taking advantage of cloud computing with this parallel intensive paradigm \citep{4839978}.
Comparisons of high-performance computing paradigms clearly show the advantages in the improvement of computation times, not only in hardware but also with differing algorithms \citep{2010NewA...15..726B}.  In the case of rendering with 3D graphics software like Blender, the choice of rendering tile size is important for optimizing
the usage of threads on the GPU for parallel processing.  Using tile sizes in powers of two is the most efficient choice for optimizing memory usage in a parallel processing scheme \citep{291528}.

\begin{figure}
\epsscale{0.8}
\plotone{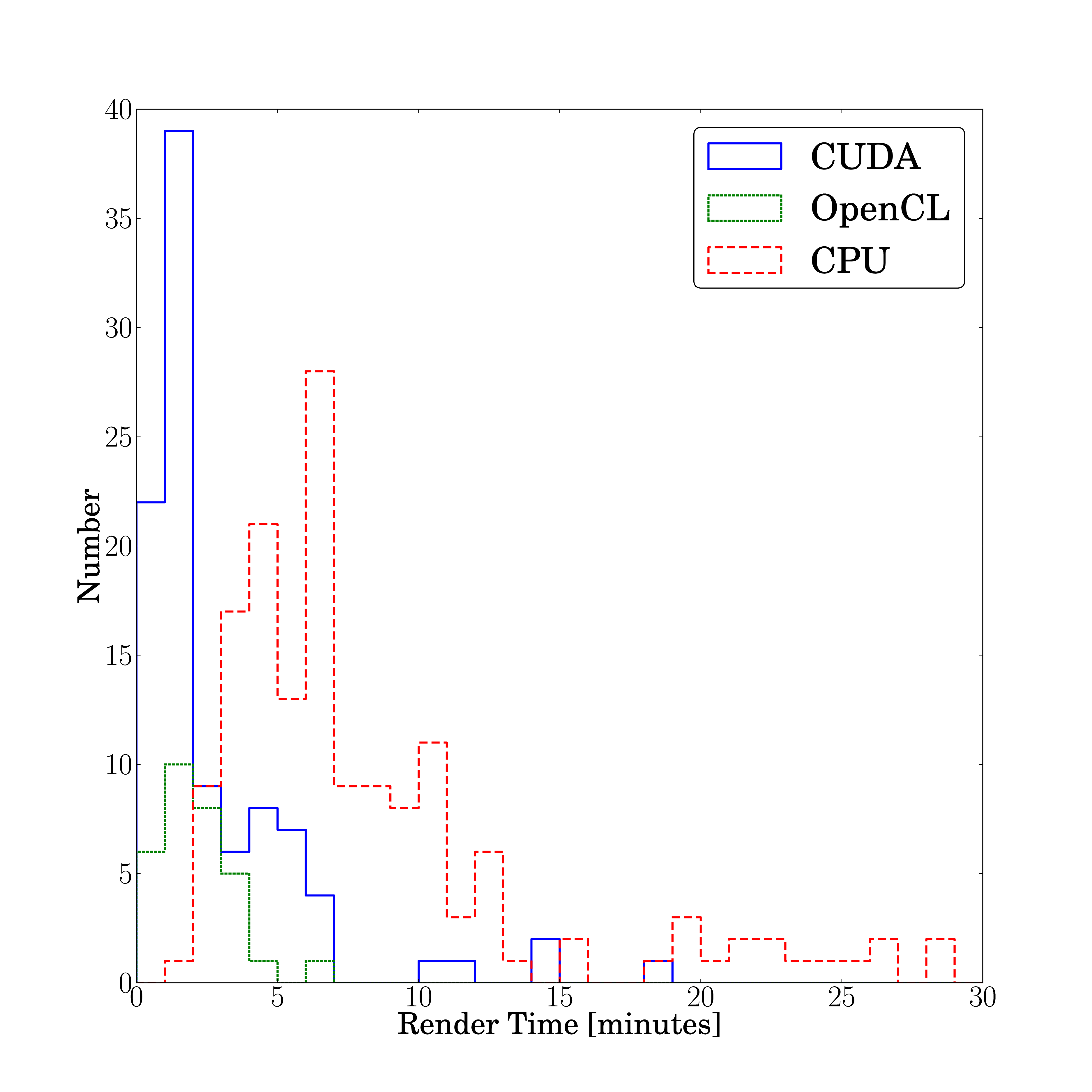}
\caption{Benchmark comparison running a Blender session in Cycles at a resolution of 960 $\times$ 540 pixels with rendering tile sizes of 8 $\times$ 8 pixels.  The distribution of render times shown are for NVidia CUDA (\textit{solid blue line}), and OpenCL (\textit{dotted green line}), and CPU processing (\textit{dashed red line}), each binned in 1 minute intervals.  The vast majority of GPU runs with CUDA and OpenCL outperform standard CPU-based rendering.\label{gputrials}}
\end{figure}

High performance computing in visualizing and animating data can be quantified with benchmarks trials comparing CPU and GPU rendering times.  A single frame benchmark comparison was completed at a resolution of 960 $\times$ 540 pixels with rendering tile sizes of 8 $\times$ 8 pixels.  The data were obtained from a variety of computer platforms and operating systems using the Cycles rendering engine (O. Amrein 2013, private communication).  The benchmark test consisted of six camera views and 12 mesh objects totaling 86,391 polygons and 91,452 vertices.  Figure \ref{gputrials} plots the distribution of times for a) NVidia CUDA, b) OpenCL, and c) CPU processing all running the same benchmark Blender file.  In these rendering trials using a GPU improves the mean rendering time by an average factor of $\sim$3.4 when compared to dual or quad-core CPU processing.  The time trial benchmarks are shown in Table \ref{gputable}.  GPU enabled rendering is advantageous when considering the number of individual frames that might contribute to an animation.

\subsection{Comparison with Other Software}
\label{comparisonsoftware}

\citet{Laramee:2008:CEC:1364835.1364838} compared a number of lower level graphics libraries, noting attributes suited to both research and industry.  The benefit of Blender for astronomers is that it is freely available open-source software that 
runs on the Linux, Mac, and Windows and can be used with the GUI or through Python scripting.  The GUI is not tied to any of the three operating systems and does not depend on the Microsoft Software Development Kit, Cocoa for Mac, or Gnome/KDE for Linux -- it looks and functions the same on every system.
Table \ref{softwarecompare} compares Blender with other 3D graphics packages.  Other packages exhibit similar capabilities in modeling and lighting, but are typically used with external rendering engines.  Blender is unique in that it provides a self-contained environment for a end-to-end workflow without the need for extra software libraries or installations.

\subsection{Required Resources}

The workflow for modeling, texture mapping, and animation lends itself to relatively modest hardware, including laptops for preview renders and basic manipulation.  However, because of the portability of the data structure, blend files can be easily copied to workstations or clusters for CPU/GPU intensive rendering output.  Blender scripting uses Python 3.2, included with the binaries and source distribution.  For the examples that are described in this paper we also make use of PYFITS and the Matplotlib libraries for data manipulation \citep{2000ASPC..216...67B, 4160265}.  In addition, a three-button mouse is recommended for moving within the 3D view space.

\section{Example Workflow Session}
\label{workflowsession}

\begin{figure}
\epsscale{0.6}
\plotone{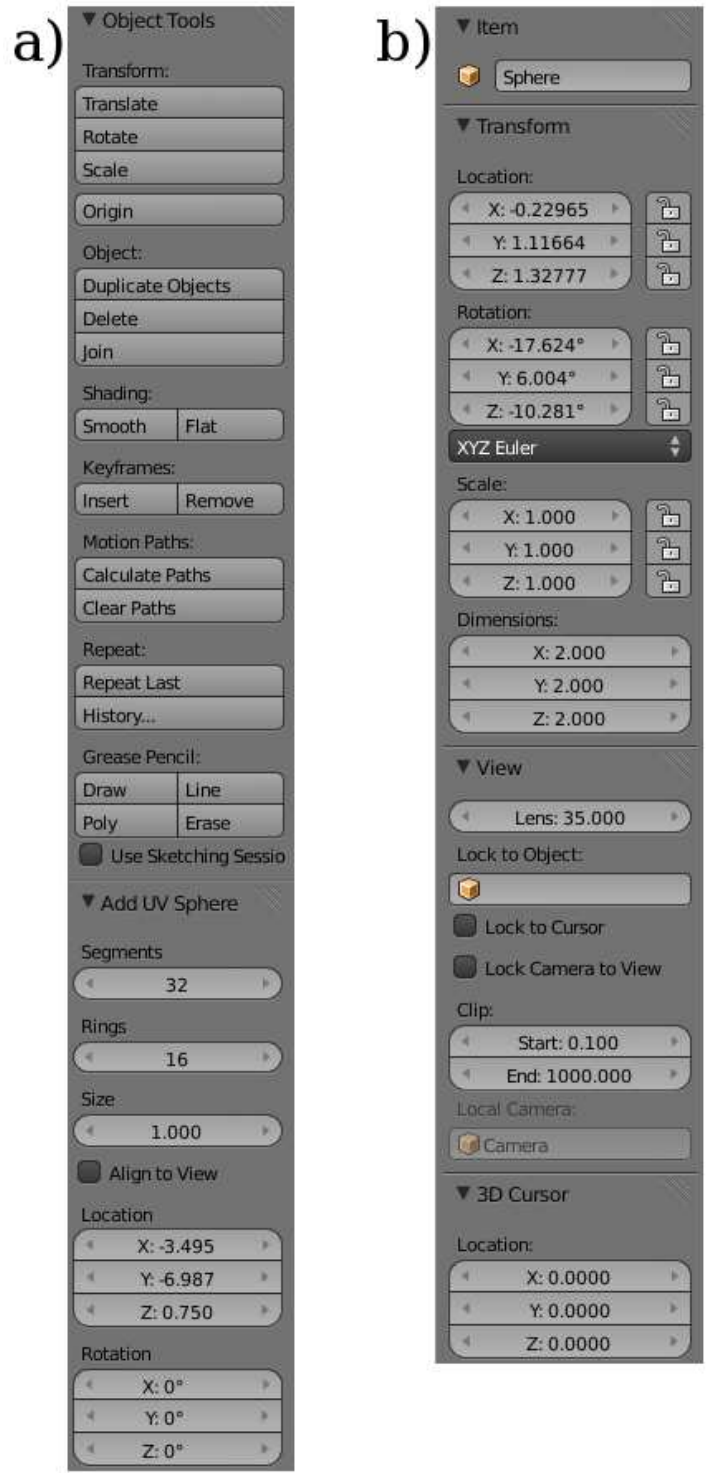}
\caption{Object tools and data transform widgets.  (\textit{a}) Object toolbar which allows the user to move, rotate, or scale a given object and keyframe any animation properties.  (\textit{b}) Transform toolbar giving the precise location, rotation, and scaling properties, camera views, and cursor properties for object placement.\label{objecttransform}}
\end{figure}

\begin{figure}
\epsscale{0.6}
\plotone{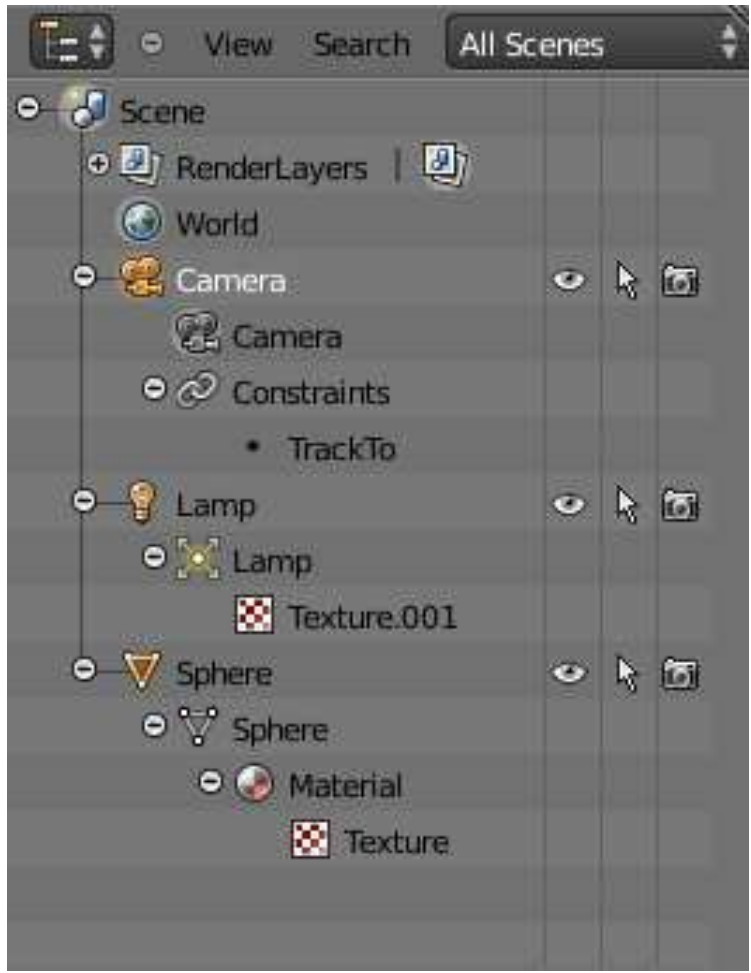}
\caption{Data outliner widget.  This hierarchical tree view depicts the program's outline of all objects within the scene.  In this example, the Camera has a "Track To" constraint applied.  The map has a simple lighting texture, and the sphere (a planet in this case) has a material and UV-mapped texture applied.  Each object is displayed in the view port, indicated by the eye icon.  Each object is also applied to at least one render layer, indicated by the camera icon on the right hand side.\label{dataoutliner}}
\end{figure}

\begin{figure}
\epsscale{1.0}
\plotone{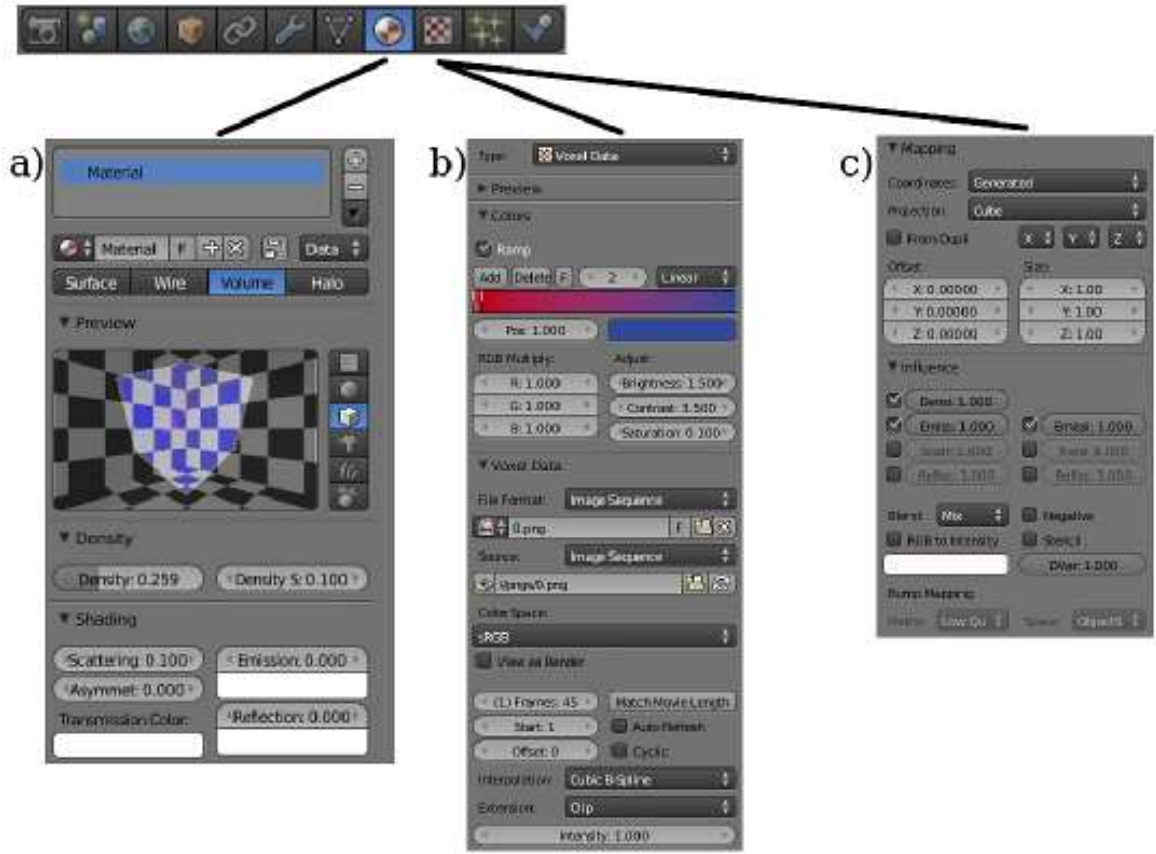}
\caption{Tab icons at top highlight the material and texture widgets.  (\textit{a}) Material widget gives the user the options to modify lighting, specular, and shading parameters, and whether what kind of material (surface, wire, volume, or halo) will be applied to the object.  (\textit{b}, \textit{c}) Texture widget for applying surface and volume textures, as well as color maps, to the objects.\label{materialstextures}}
\end{figure}

\begin{figure}
\epsscale{1.0}
\plotone{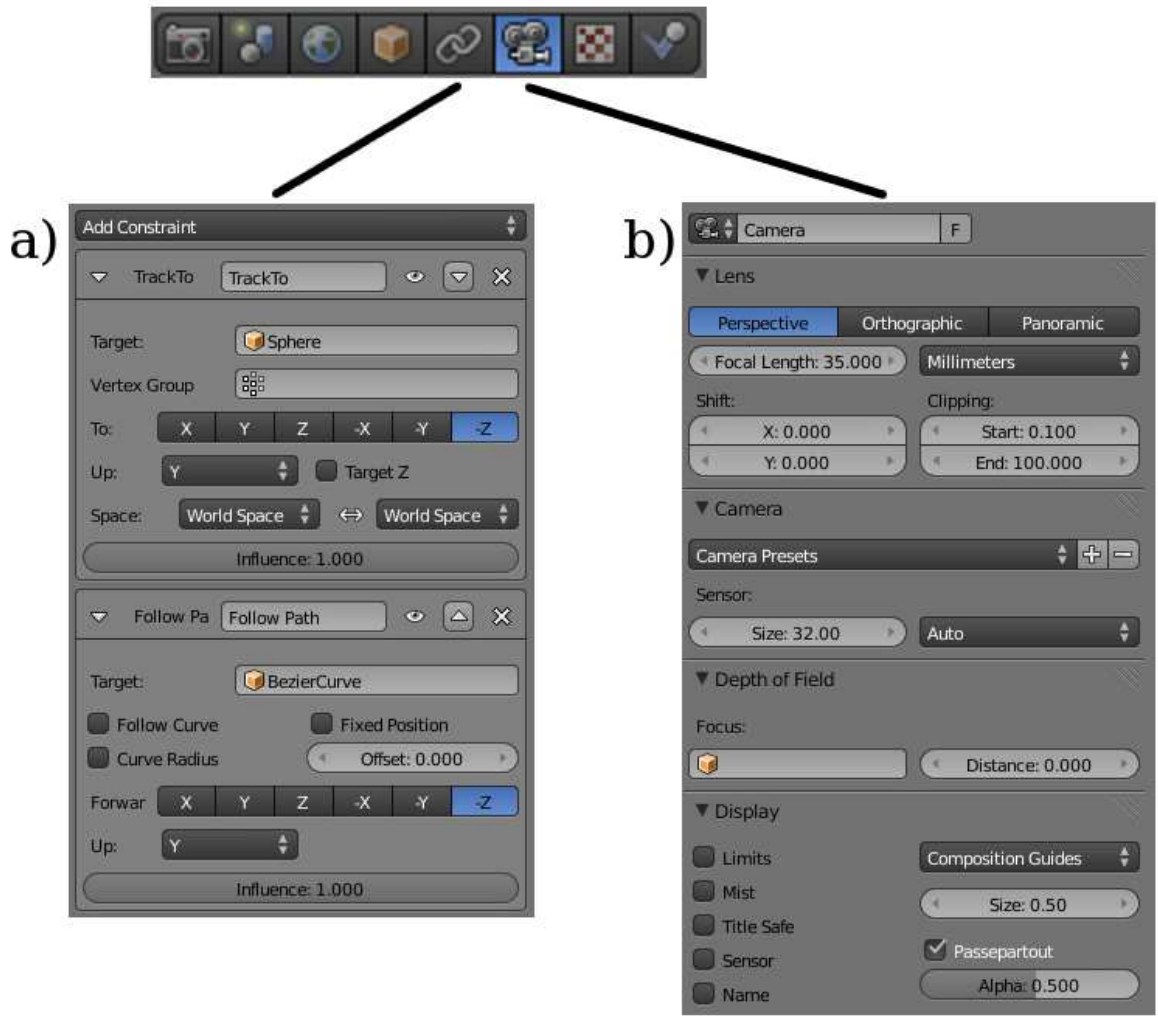}
\caption{Tab icons at top highlight the camera constraint and properties widgets.  (\textit{a}) Examples of "Track To" and "Follow Path" constraints.  The objects local axes are oriented so that the camera will follow the path of a B{\'e}zier curve while pointing at a UV-Sphere mesh.  (\textit{b}) Use of a 35 mm lens in a perspective projection as well the clipping limits of the camera.  This clipping parameter controls what can be seen by the camera in the animation scene. \label{camera}}
\end{figure}

We describe a workflow session applicable to visualizing a data cube in 3D.  The motivation
behind this example is to put into practice the concepts outlined in section \ref{blender}
and how they relate to actual astronomical data.  The example position-position-frequency FITS data cube
is from the M81 dwarf A galaxy observed with the The {\ion{H}{1}} Nearby Galaxy Survey \citep[THINGS;][]{2008AJ....136.2563W}.  Studying the data cube in 3D allows for the understanding of its dynamical structure more clearly than with examination of 2D position-velocity diagrams.  The naturally weighted cube is cropped to 130 pixels in Right Ascension by 130 pixels in Declination by 45 pixels in frequency space.  A script for this example will be provided in the summary section.  For research purposes, it is meant for visual inspection of a data cube and for seeing dynamic features of a galaxy that can be difficult to visualize in 2D.  However, the general application of viewing a 3D data cube also has uses for public outreach -- explaining what astronomical data cubes look like, or,
in the case of a simulation data cube, moving the camera around during an animation.  The same functionality in Blender applies in both cases.  A number of key operational details are relevant for any session.  Keyboard shortcuts rapidly increase productivity in Blender; we refer the reader to the Blender documentation for comprehensive listings.  All commands listed work the same regardless of the operating system.  We will refer to elements of the GUI as dialogs, enlarged and indicated for clarity in Figures \ref{objecttransform}--\ref{camera}.  
The guidelines and figures outlined here serve as a starting point for the reader; as with any visualization package practice and experience will be required for optimal output.  We denote menu items in italics as \textit{Menu, Submenu, Item} and 
button commands, dialogs, and widgets in \textbf{Bold}.

\begin{figure}
\epsscale{1.0}
\plotone{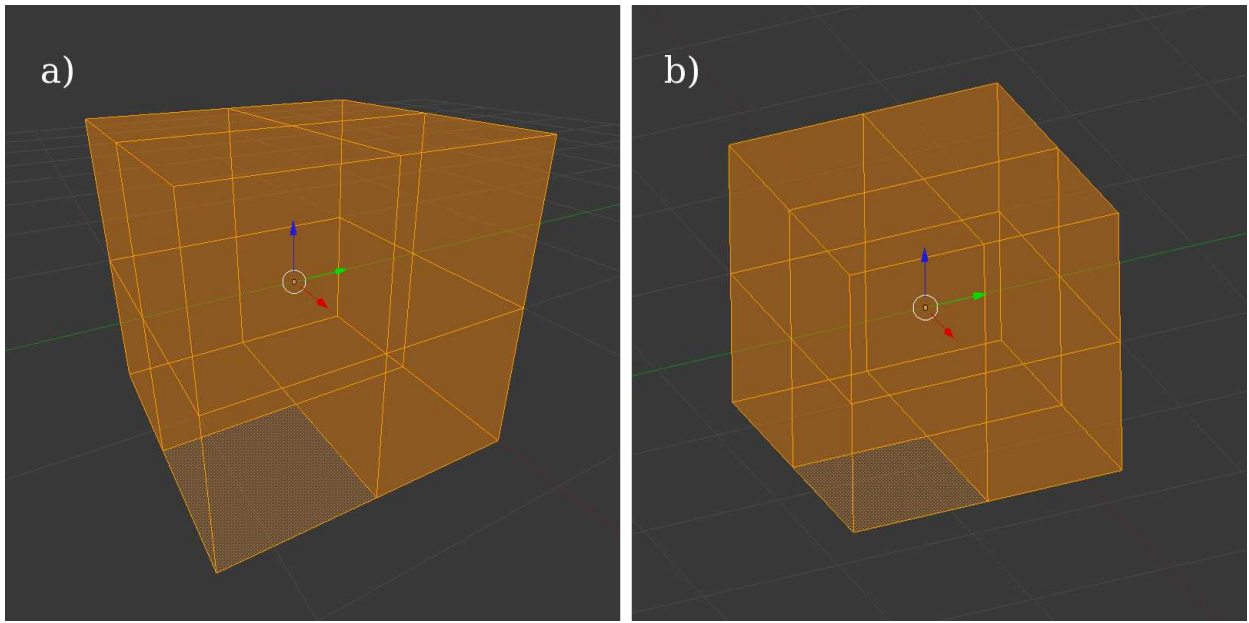}
\caption{Differences between (\textit{a}) perspective and (\textit{b}) orthographic projection.  Each view has applications in astronomical visualization.\label{orthoperspec}}
\end{figure}

\begin{itemize}

    \item  New mesh data objects are inserted into the viewspace (called a \textbf{Scene}) at the location of the 3D cursor.  The cursor's location is shown by the red and white dashed circle and black cross-hairs (see Figure~\ref{controls}).  This can be moved with primary (for most users usually left) mouse button or reset to the origin in the right hand side \textbf{transformation toolbar}.

    \item  A useful workspace can be utilized by accessing the menu \textit{View, Toggle Quad View}.  This shows the top, front, and right side views as well as a preview of what the camera will see in the final animation.
    
    \item  For draft animations, we recommend using the HD 1280x720 pixel preset in the right hand side \textbf{Dimensions} dialog, with a NTSC standard frame rate of 29.97 frames per second\footnote{ITU Recommendation BT.709: http://www.itu.int/rec/R-REC-BT.709/en}.  Once the reader is satisfied with the low resolution results higher resolutions can be rendered if desired.

    \item  Objects are selected with the right mouse button.  Objects can be edited by pressing the TAB key or choosing the \textbf{Mode} from the lower left hand drop down menu.
    
    \item  Blender supports both orthographic and perspective projection.  This can be selected with under the \textit{View, View Persp/Ortho} option and also changed in the camera tab (Figure~\ref{orthoperspec}).
    
    \item  The scale setting in the \textbf{transformation toolbar} allows units to be scaled to the data being visualized in the scene, whether they are pixels, channels, astronomical units, or kiloparsecs.

\end{itemize}

\subsection{Data Model and Volume Textures}

Preparation of data cubes can be accomplished with PYFITS or CFITSIO\footnote{http://heasarc.gsfc.nasa.gov/fitsio/}, and saving each image with Matplotlib.  Any Hanning smoothing should be completed prior to graphics export.  Files should be named sequentially, e.g., channel0001.jpg, channel0002.jpg, channel0003.jpg, etc., and placed in their own subdirectory.  The dynamic range of the image will be clipped by this procedure.  However, this does not affect the output since the Python script controlling the Blender file can modify the scaling before rendering.

Data cubes in 3D space can be represented by the simplest of mesh primitives -- a cube.  Cubes can be inserted by accessing the menu items \textit{Add, Shape, Cube}.  Different data objects can be selected with the \textbf{Data Outliner} (Figure~\ref{dataoutliner}).  We can then modify the mesh material options by clicking the red globe icon for the \textbf{Materials} widget (Figure~\ref{materialstextures}).  We choose a volumetric material, set the graphic density to 0.0, density scale to 2.0, scattering to 1.4, and reflection to 0.0.

We then choose the \textbf{Texture} widget where the data can be added as a set of image planes.  A new texture can be set to import \textbf{Voxel Data} under the \textbf{Type} dropdown dialog.  Working from the top of this dialog, the color map \textbf{Ramp} dialog should be checked under color choices to set the color scale.  The color scheme choice is important in visualization \citep{Rhyne:2012:ACT:2343483.2343484}, and Blender itself can be used to understand color spaces in 3D\footnote{http://www.photo-mark.com/notes/2013/mar/13/color-theory-blender/}.  It is best to choose contrasting colors to distinguish the signal from the noise \citep{2007AJ....133..598R}.  The main \textbf{Voxel Data} drop-down menu should be changed to \textbf{Image Sequence} and the first image should be selected.  Blender will automatically load the rest of the files in the directory sequentially.
Below the \textbf{File} dialog, the start and stop frames (channels in this case) can be chosen.

The mapping of the cube needs to be set to \textbf{Generated}, with a projection of \textbf{Cube}.  Under the \textbf{Influence} dialog, \textbf{Density}, \textbf{Emission}, and \textbf{Emission Color} should all be selected.  The \textbf{Blend} method should be set to \textbf{Mix}.

\subsection{Camera Setup}

The camera can be selected by right clicking the pyramid shaped object in the viewport, followed by choosing the \textbf{Camera} tab (Figure~\ref{camera}).  At this stage the user can choose the lens configuration with perspective or orthographic and focal length.  Different types of visualization will require different viewing perspectives or, perhaps, multiple ones for split screen shots \citep{Carlbom:1978:PGP:356744.356750}.  The camera depth of field and sensor size can also be set with composition guides for alignment.

To the left of the \textbf{Camera} tab is small chain icon for adding constraints to the camera view during the animation.  For example the \textbf{Track To} feature allows the center of the camera's field of view to always point to a particular object mesh during the animation.  The  \textbf{Follow Path} option is also particularly useful, as the camera can be locked to follow a predetermined path.

\begin{figure}
\epsscale{1.0}
\plotone{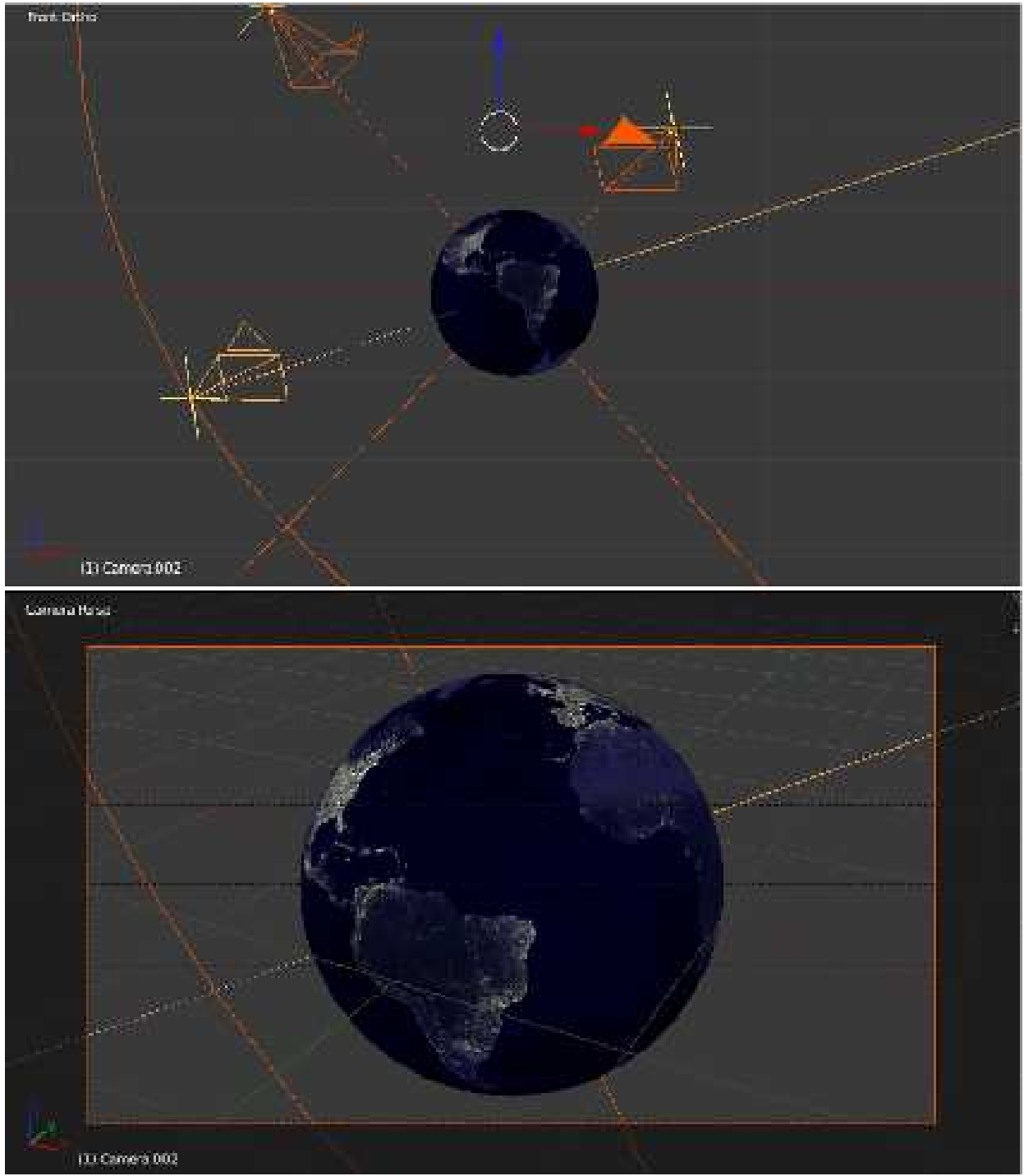}
\caption{Multiple camera angles are shown, with vectors pointing at an Earth surface mesh that are normal to the film planes (\textit{yellow crosses}).  The camera in the lower left of the \textit{top panel} is following a B{\'e}zier curve path while tracking the planet model.  The \textit{lower panel}
shows a perspective projection through the camera view.\label{cameraduplicate}}
\end{figure}

It is also useful in any animation sequence to have multiple camera angles - the camera can be duplicated with the \textbf{Duplicate Objects} button on the far left hand side of the interface.  This will copy the camera, its constraint parameters, and allow the user to freely move the new camera object to a different location.   A schematic setup of these features and camera view is shown in Figure~\ref{cameraduplicate}.

\subsection{Animation and Keyframes}

\begin{figure}
\epsscale{1.0}
\plotone{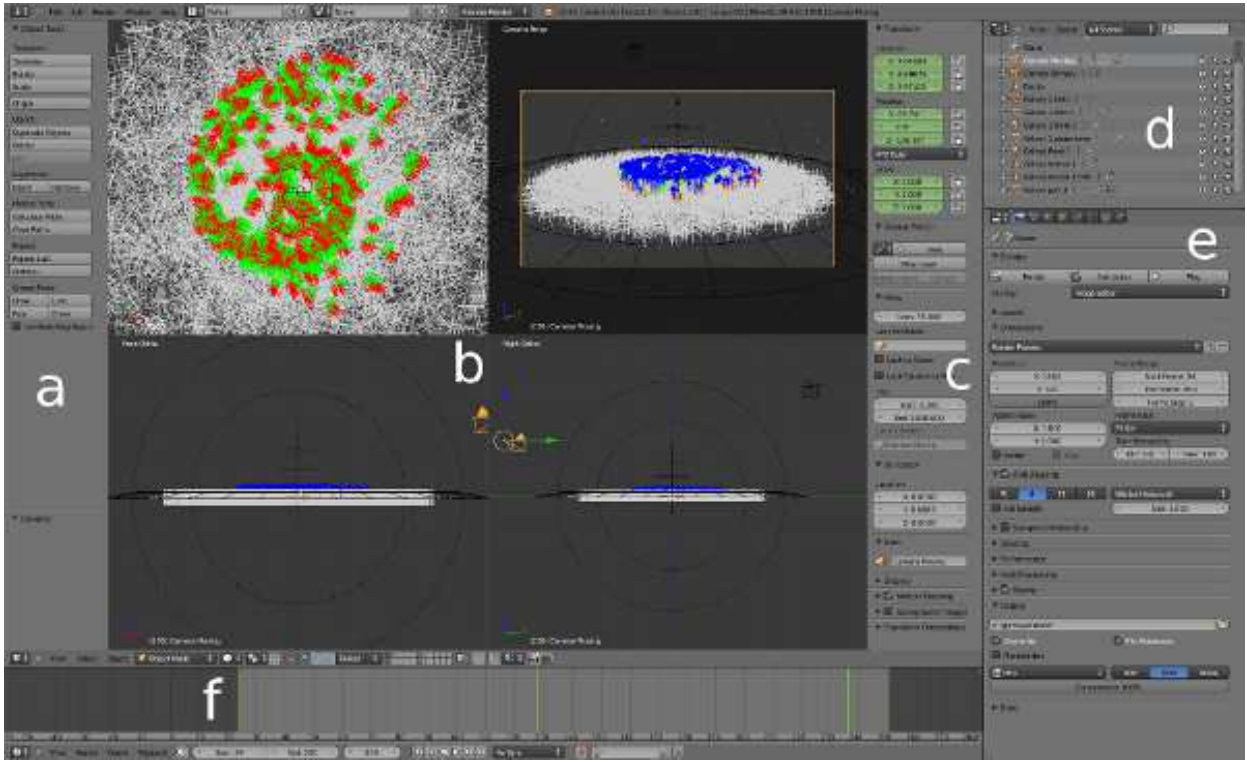}
\caption{Main Blender interface.  (\textit{a}) Object Toolbar for manipulating a selected object.  (\textit{b}) Main 3D view port.  In this example a "Quad View" is shown for the top, front, and right orthographic perspectives, as well as the preview of the camera angle.  (\textit{c}) Transformation toolbar which allows precise control of objects.  (\textit{d})  Hierarchical data outliner, summarizing the properties and settings of each data structure.  (\textit{e}) Properties panel for the camera, world scene environment, constraints, materials, and textures.  (\textit{f}) Animation time line, frames in the video animation, and yellow marks indicating keyframes for selected objects.\label{maininterface}}
\end{figure}

The animation tool and timeline is located at the bottom of the Blender interface (Figure~\ref{maininterface}).  It contains a scrolling timeline, with a green vertical line indicating the current frame, controls for playing the animation, and fast forwarding and rewinding (similar to any media player).  The animation timeline defaults to frame one.  Once the mesh cube object (our data container) has been loaded with the volumetric texture, it can be positioned and animated.  We can approach this by one of two methods -- one animates the data cube itself, and the other keeps the data cube static while moving the camera on a fixed track.  We create a 20 s animation at approximately 30 frames per s$^{-1}$, for a total of 600 frames.  Therefore, the \textbf{End Frame} should be set to 600. Our data cube is aligned with right ascension along the \textit{x}-axis (\textit{red}), declination along the \textit{y}-axis (\textit{green}), and frequency along the \textit{z}-axis (\textit{blue}).

\textit{\textbf{Rotating the data cube.}}  We can keyframe the initial state by pressing the \textbf{Insert Keyframe} button on the \textbf{Object Tools} panel (Figure~\ref{objecttransform}) and choosing the option \textbf{Rotation}, which locks the rotation of the mesh cube for that particular frame.  The cube can be rotated about the \textit{z}-axis (\textit{blue}) by moving the green marker on the animation timeline to frame 300 (halfway through the planned animation) and changing the \textit{z}-rotation (right hand \textbf{transformation toolbar}) to 180 degrees and again choosing \textbf{Insert Keyframe, Rotation}.  This is repeated a third and final time for frame 600, with the \textit{z}-rotation to 360 degrees.  The play button can then be engaged on the bottom animation toolbar and a preview of the animation motion will be shown.

\textit{\textbf{Revolving the camera.}}  In this example we keep the cube mesh object stationary and move the camera along a path.  This can be accomplished by choosing the \textit{Add, Curve, Circle} at the top of the main interface
(see Figure~\ref{maininterface}) and scaling it (\textbf{s}~key) to the size of the path needed to view the data cube.  The path animation can be keyframed under the \textbf{Object Data} tab just like any other object in Blender, rotating the path 360 degrees (or more) during the course of our 600 frame, 20 s output.  The camera can then be constrained with a \textbf{Follow Path} constraint with its local \textit{y}-axis normal to the circle, and further constrained with a \textbf{Track To} control with the \textit{z}-axis of the camera normal to the focal plane pointing to the data cube (Figure~\ref{camera}).

\subsection{Rendering and Output}

\begin{figure}
\epsscale{1.0}
\plotone{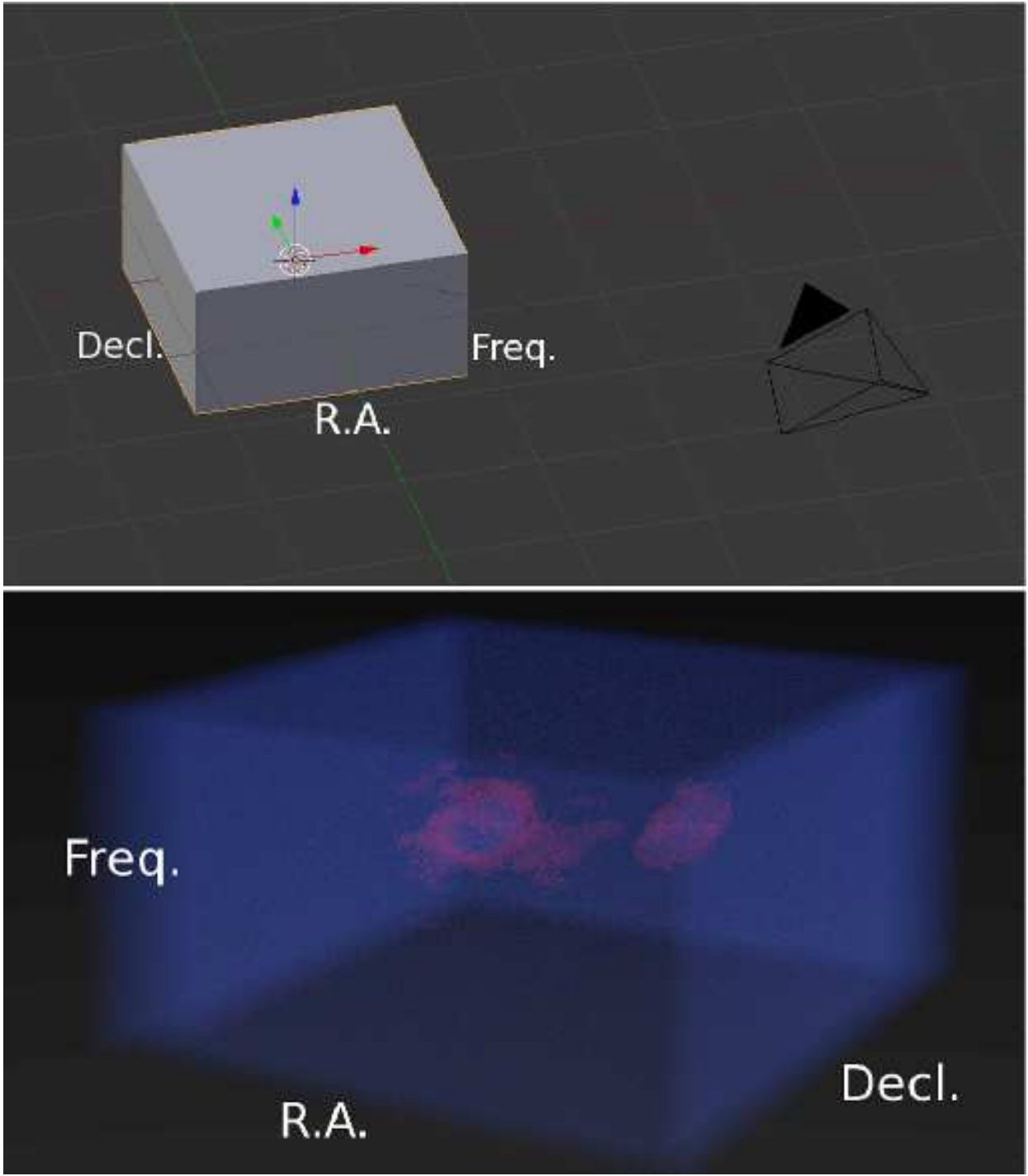}
\caption{\textit{Top}: View space setup for a data cube and camera in an orthographic projection.  \textit{Bottom}: Final perspective projection rendering of an {\ion{H}{1}} data cube.  \label{datacube}}
\end{figure}

Once the data cube, tracking paths, and camera have been keyframed, the output can be rendered.  Under the render tab there are two options: \textbf{Render} and \textbf{Animation}.  \textbf{Render} will allow the user to view the output of the current frame as viewed in the animation timeline.  \textbf{Animation} will render out all the frames of the current scene.  Most of the default parameters will work well for typical visualization scenarios.  We recommend checking the \textbf{Stamp} box which will place metadata about the animation on each image, including parameters used in the animation settings.  Output can be specified as still images for each frame or one of several video outputs.
The setup and a rendered frame from the final output is depicted in Figure~\ref{datacube}.

\section{Example Visualizations}
\label{examples}

In addition to the data cube example highlighted in section~\ref{workflowsession}, we outline a number of other astronomical visualizations that are likely of use to the community.  Blender has found successful use in materials science, medicine, biology, fluid dynamics, and network theory \citep{Norman2012, 2013JPhCS.410a2169C, Andrei2012, 2011PhFl...23i1109R, 2005PhRvE..72b6107V}.  In the case of studying large scale structure, Blender provided an interface for examining redshift distortions and correlations between quasar absorption line systems and luminous red galaxies \citep{1367-2630-10-12-125015}.  Results from these kinds of studies are extremely challenging to discern with two-dimensional plots.  As larger catalogs with expanded parameter sets begin to become the norm rather than the exception, having a 3D graphics and rendering environment like Blender for astronomers will prove to be an invaluable tool.
We provide the following examples to illustrate starting points for those interested in pursuing
their own visualization scenarios for both exploratory research as well as outreach endeavors.  The visualizations are designed to put into practice aspects of the interface described in section \ref{blender}.  Example data, files, and tutorials will be provided for download in section \ref{summary}.

\subsection{Astronomical Catalog}

\begin{figure}
\epsscale{1.0}
\plotone{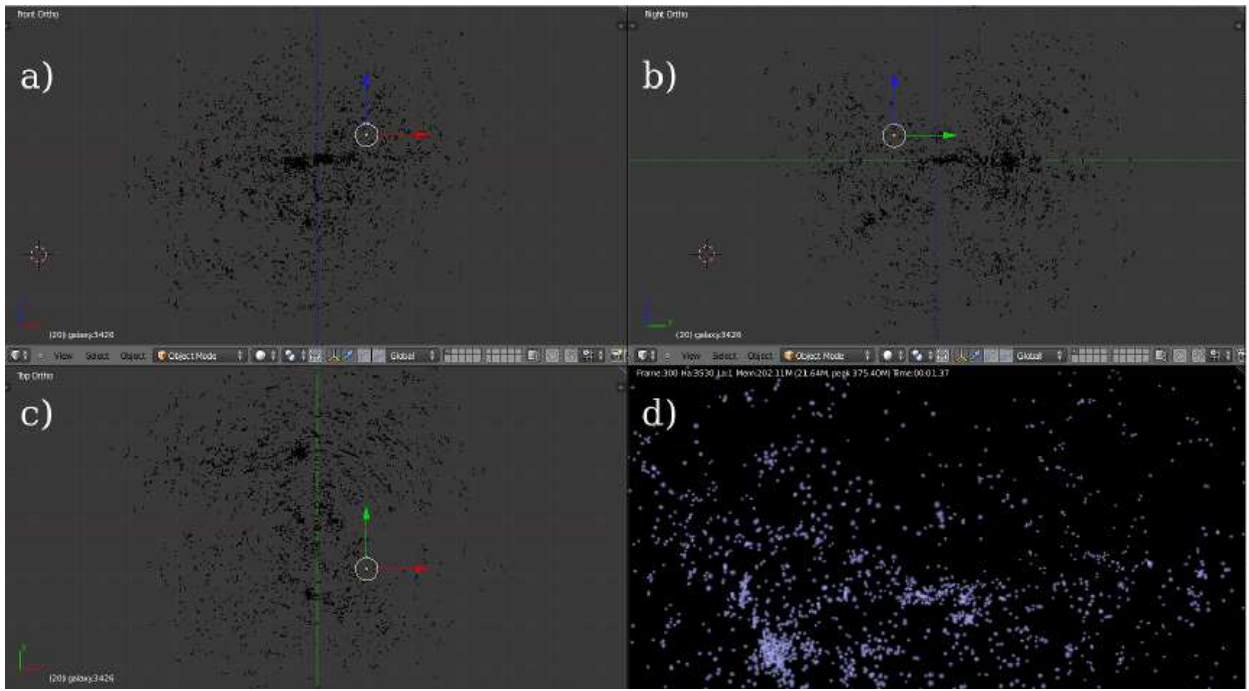}
\caption{Three-dimensional view of a nearby galaxy catalog ($cz_{\odot}<$ 3000 km~s$^{-1}$) from the Extragalactic Distance Database (EDD).  (\textit{a}, \textit{b}, \textit{c}) \textit{Front}, \textit{right}, and \textit{top} orthographic projections, respectively. (\textit{d}) Single render frame from the animation in a perspective projection.\label{catalog}}
\end{figure}

We can use a small sample of galaxy distances to illustrate the mapping and animating of a catalog fly-through.  Figure~\ref{catalog} shows a Blender scene setup and a single frame render of data from the Extragalactic Distance Database \citep[EDD\footnote{Data available at http://edd.ifa.hawaii.edu/};][]{2009AJ....138..323T} of galaxies in the nearby Universe ($cz_{\odot}<$ 3000 km~s$^{-1}$).

This example can utilized in scientific data exploration because we can examine the overall structure
of the catalog in real time with the 3D view space.  Since each galaxy has a unique object identifier in the data structure, we can use the Blender search function (\textbf{Spacebar}) and period key (\textbf{.}) to immediately center and zoom to any galaxy.  This is useful for research, but also for planning camera movements for rendering if so desired.

Importing and rendering an astronomical catalog makes use of the simplest type of mesh - a vertex point, Python scripting, texturing, animation, camera movement, and rendering.  ASCII catalogs can be read into Python lists.  A template object can be created
with a single vertex textured with a halo in the \textbf{Materials} tab, size set to 0.020.
Under the \textbf{Textures} tab, the \textbf{Type} should be set to "\textit{Blend}."  The camera can
then be keyframed for animation.

\subsection{\textit{N}-body Simulation}

\begin{figure}
\epsscale{1.0}
\plotone{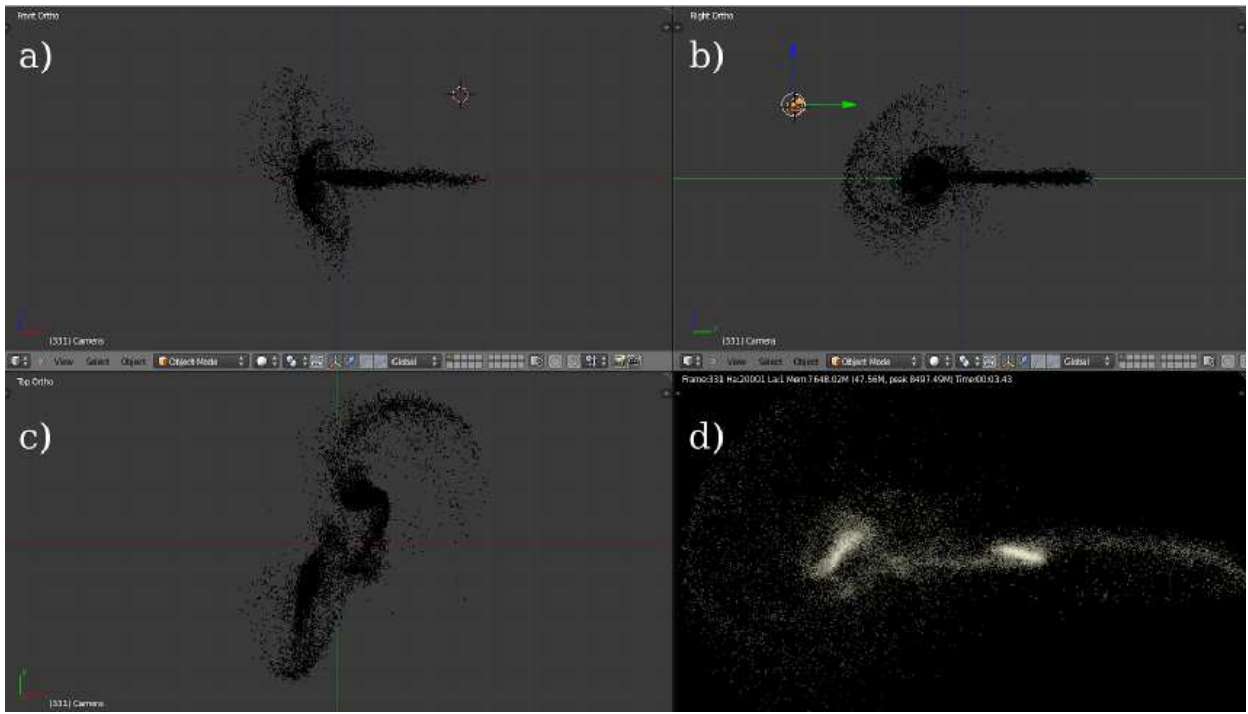}
\caption{Three-dimensional view of a simulation with colliding galaxies.  (\textit{a}, \textit{b}, \textit{c}) \textit{Front}, \textit{right}, and \textit{top} orthographic projections, respectively. (\textit{d}) Single render frame from the animation in a perspective projection.\label{nbody}}
\end{figure}

This example of a galaxy simulation was generated with GADGET-2~\citep{2005MNRAS.364.1105S}.  Each large spiral disk galaxy has 10000 disk particles and 20000 halo particles with Milky Way scale lengths and masses.  The simulation is run for approximately 1100 timesteps for a total simulation runtime of 2 billion yr.   We read in the particle \textit{x}, \textit{y}, \textit{z} coordinates as single vertex with a small Gaussian halo as its materials texture.  The snapshot file for each time step is keyframed as one frame in the animation.  In addition, a B{\'e}zier curve can be added to the scene as an object path \citep{Farouki:2012:BPB:2221977.2222160}.  The camera can then be flown along the curve as the galaxy interaction progresses (Figure~\ref{nbody}).

Blender has excellent utility for research purposes in rendering simulations.  
Many simple astronomical simulations are of the format \textit{x-y-z} over time
and are easily accommodated.
The camera control in particular is useful for moving through 3D space as simulation
timesteps play through an animated rendering.  The procedure is very similar
to opening a catalog in the previous example; for a simulation we can open different snapshots
and keyframe to create a smooth animation.  Snapshot files can be read into a Python list or dictionary.  A template object can be created
with a single vertex textured with a halo in the \textbf{Materials} tab, size set to 0.020.
Under the \textbf{Textures} tab, the \textbf{Type} should be set to "\textit{Blend}."  The positions of each
particle can then be loaded with the Python API. The camera can be keyframed for animation
and a final rendering can be generated.

\subsection{Asteroid Models}

\begin{figure}
\epsscale{1.0}
\plotone{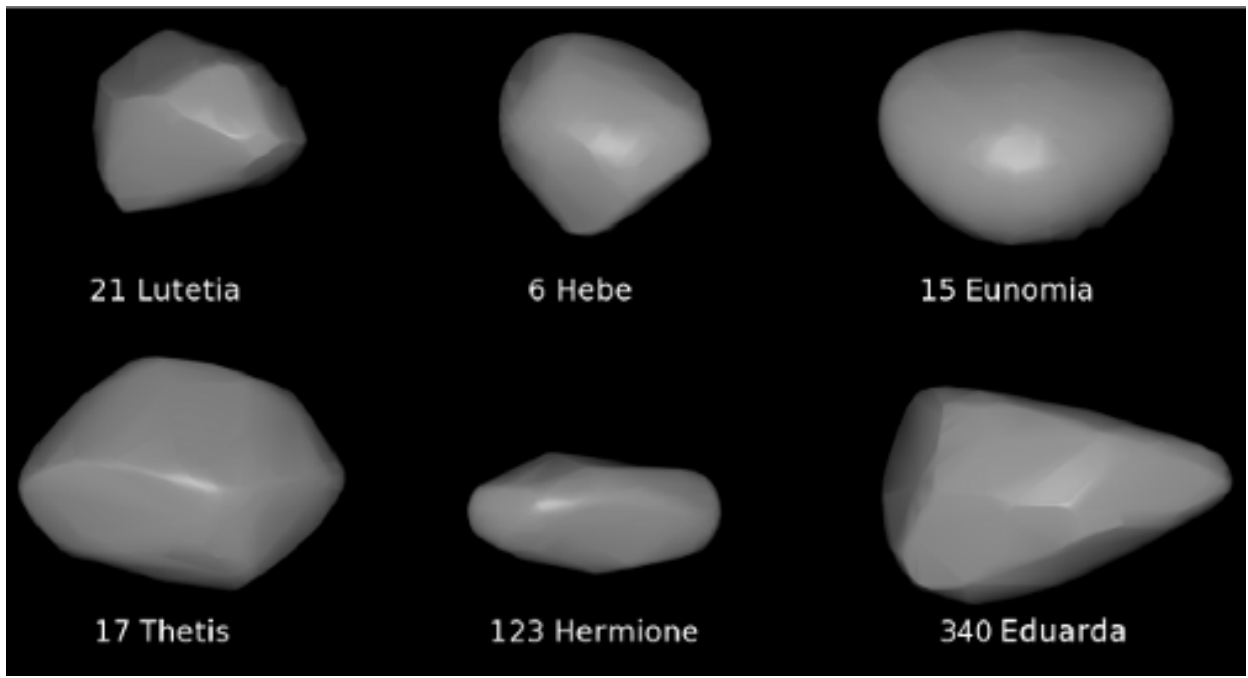}
\caption{Asteroid models in this figure show how OBJ files can be rendered in Blender.
The models are not shown to relative scale and have been increased in size for detail.
\label{asteroids}}
\end{figure}

We show a number of 3D asteroid models based on data available from the Database for Asteroid Models from Inversion Techniques \citep{2010A&A...513A..46D}.  The database provides Wavefront object files of each asteroid model, which can easily be loaded into the 3D viewport.  For this example, we want uniform lighting to be able to see the entire surface model as they are rotated.  This can be accomplished by turning on Environment Lighting in the \textbf{World} tab (small globe icon, see Figure~\ref{camera}).
Figure~\ref{asteroids} exhibits an example of an orthographic projection and shows a sample of six asteroids, enlarged to show the smoothing texture on the surfaces.

This example exhibits how 3D mesh models in OBJ files can be imported into Blender
for animation and rendering.  OBJ files from data such as our asteroid example can be used for research purposes, but also have merit in video renderings for public outreach.  This can be accomplished in the upper left corner of the GUI and choosing \textit{File, Import, Wavefront (*.obj)}.  The object can be scaled in the \textbf{Transform} dialog on the right side of the GUI, or by pressing the 'S' key.  The asteroid mesh object can be selected with the right mouse button,
keyframed for animation with the "I" key, and rotated with the "R" key.


\section{Summary}
\label{summary}

The availability of hardware and software that allows the creation of high-quality
3D animations for astronomy brings a vivid new way for astronomers to visualize
and present scientific results.  Animations are an important part of the astronomical
toolset that can show expanded phase spaces, the time domain, or multidimensional tables
and catalogs beyond the limitations posed by two-dimensional plots and images.

An introduction has been presented showing the software program Blender as a 3D animation and rendering package for astronomical visualization.  We have reviewed the importance of visualization in the scientific disciplines, with examples from astronomy.  Features of the program include GPU processing, a variety of rendering engines, and an API for scripting.  An overview of the principles
in the animation workflow include modeling, volumetric and surface texturing, lighting,
animation, camera work, rendering, and compositing.  With these general steps in mind
various astronomical datasets -- images, data cubes, and catalogs can be imported 
and visualized.

An example workflow session with an astronomical data cube that shows the structure and dynamics of a small dwarf galaxy has been presented,
giving settings and recommendations volumetric texturing of a cube model, animation,
and animation output.  A number of other example are shown from different areas
of astronomy to give a broad scope of possibilities that might exist with astronomical data.

A demonstration video as well as Blender files, Python scripts, and basic tutorials of the principles and examples outline in this paper are available at:\\ http://www.cv.nrao.edu/$\sim$bkent/computing/kentPASP.html


\begin{thebibliography}{99}


\bibitem[{Ament {et~al.}(2010)Ament, Weiskopf, \&
  Carr}]{Direct_Interval_Volume_Visualization}
Ament, M., Weiskopf, D., \& Carr, H. 2010, IEEE Trans. Visual. Comput. Graph., 16, 1505

\bibitem[{{Andrei} {et~al.}(2012)}]{Andrei2012}
Andrei, R.~M., Callieri, M., Zini, M., Loni, T., Maraziti, G., Pan, M., \& Zopp\'{e}, M. 2012, BMC Bioinf., 13(Suppl 4), S16

\bibitem[{Autin {et~al.}(2012)Autin, Johnson, Hake, Olson, \&
  Sanner}]{Autin:2012:UUC:2412364.2412538}
Autin, L., Johnson, G., Hake, J., Olson, A., \& Sanner, M. 2012, IEEE Comput.
  Graph. Appl., 32, 50

\bibitem[{Balakrishnan \&
  Kurtenbach(1999)}]{Balakrishnan:1999:EBC:302979.302991}
Balakrishnan, R., \& Kurtenbach, G. 1999, in Proc. of the SIGCHI
  Conf. on Human Factors in Computing Systems (New York: ACM), 56

\bibitem[{{Barrett} \& {Bridgman}(2000)}]{2000ASPC..216...67B}
{Barrett}, P.~E., \& {Bridgman}, W.~T. 2000, in ASP Conf. Ser. 216, Astronomical Data Analysis Software and
  Systems IX, ed. N.~{Manset}, C.~{Veillet}, \& D.~{Crabtree} (San Francisco: ASP), 67

\bibitem[{{Bate} {et~al.}(2010){Bate}, {Fluke}, {Barsdell}, {Garsden}, \&
  {Lewis}}]{2010NewA...15..726B}
{Bate}, N.~F., {Fluke}, C.~J., {Barsdell}, B.~R., {Garsden}, H., \& {Lewis},
  G.~F. 2010, NewA, 15, 726

\bibitem[{{Bergeron} \& {Foulks}(2006)}]{2006ASPC..359..285B}
{Bergeron}, R.~D., \& {Foulks}, A. 2006, in ASP Conf. Ser. 359, Numerical Modeling of Space Plasma Flows, ed.
  G.~P. {Zank} \& N.~V. {Pogorelov} (San Francisco: ASP), 285

\bibitem[{{Bertin} \& {Arnouts}(1996)}]{1996A&AS..117..393B}
{Bertin}, E., \& {Arnouts}, S. 1996, A\&AS, 117, 393

\bibitem[{Birn(2000)}]{Birn:2000:DLR:556898}
Birn, J. 2000, Digital Lighting and Rendering (second ed.; Thousand Oaks: New Riders)

\bibitem[{Blinn(1978)}]{Blinn:1978:SWS:965139.507101}
Blinn, J.~F. 1978, SIGGRAPH Comput. Graph., 12, 286

\bibitem[Boch et al.(2011)]{2011ASPC..442..683B} Boch, T., Oberto, A., 
Fernique, P., \& Bonnarel, F.\ 2011, in ASP Conf. Ser. 442, Astronomical Data Analysis Software and Systems XX, ed. I.~N. Evans, A. Accomazzi, D.~J. Mink, \& A.~H. Rots (San Francisco: ASP), 683 

\bibitem[{{Borkin} {et~al.}(2007){Borkin}, {Goodman}, {Halle}, \&
  {Alan}}]{2007ASPC..376..621B}
{Borkin}, M., {Goodman}, A., {Halle}, M., \& {Alan}, D. 2007, in ASP Conf. Ser. 376, Astronomical Data
  Analysis Software and Systems XVI, ed. R.~A. {Shaw}, F.~{Hill}, \& D.~J.
  {Bell} (San Francisco: ASP), 621

\bibitem[{{Brinkmann}(2008)}]{Brinkmann2008}
Brinkmann, R. 2008, The Art and Science of Digital Compositing: Techniques for Visual Effects, Animation and Motion Graphics (second ed.; Burlington: Morgan Kaufmann)

\bibitem[{Buddelmeijer \& Valentijn(2013)}]{Buddelmeijer}
Buddelmeijer, H., \& Valentijn, E. 2013, Exp. Astron., 35, 283

\bibitem[{Carlbom \& Paciorek(1978)}]{Carlbom:1978:PGP:356744.356750}
Carlbom, I., \& Paciorek, J. 1978, ACM Comput. Surv., 10, 465

\bibitem[Cazotto et al.(2013)]{2013JPhCS.410a2169C} Cazotto, J.~A., Neves, 
L.~A., Machado, J.~M., et al.\ 2013, Journal of Physics Conference Series, 
410, 012169 

\bibitem[{{Comparato} {et~al.}(2007){Comparato}, {Becciani}, {Costa},
  {Larsson}, {Garilli}, {Gheller}, \& {Taylor}}]{2007PASP..119..898C}
{Comparato}, M., {Becciani}, U., {Costa}, A., {Larsson}, B., {Garilli}, B., {Gheller}, C., \& {Taylor}, J. 2007, \pasp, 119, 898

\bibitem[{{Crassin} {et~al.}(2009)}]{Crassin2009}
{Crassin}, C., {Neyret}, F., {Lefebvre}, S., {Eisemann}, E. 2009,
in Proc. 2009 Symp. on Interactive 3D Graphics and Games (New York: ACM), 15

\bibitem[{Debevec(1998)}]{Debevec:1998:RSO:280814.280864}
Debevec, P. 1998, in Proc. twenty-fifth Annual Conference on Computer
  Graphics and Interactive Techniques (New York: ACM), 189

\bibitem[{Diehl(2007)}]{Diehl:2007:SVV:1209814}
Diehl, S. 2007, Software Visualization: Visualizing the Structure, Behaviour,
  and Evolution of Software (Heidelberg: Springer-Verlag)

\bibitem[{{Draper} {et~al.}(2008){Draper}, {Berry}, {Jenness}, {Economou}, \&
  {Currie}}]{2008ASPC..394..339D}
{Draper}, P.~W., {Berry}, D.~S., {Jenness}, T., {Economou}, F., \& {Currie},
  M.~J. 2008, in ASP Conf. Ser. 394, Astronomical Data Analysis Software and Systems XVII, ed. R.~W.
  {Argyle}, P.~S. {Bunclark}, \& J.~R. {Lewis} (San Francisco: ASP), 339

\bibitem[{Drebin {et~al.}(1988)Drebin, Carpenter, \&
  Hanrahan}]{Drebin:1988:VR:378456.378484}
Drebin, R.~A., Carpenter, L., \& Hanrahan, P. 1988, SIGGRAPH Comput. Graph.,
  22, 65

\bibitem[{Duff(1985)}]{duff1985}
Duff, T., 1985, SIGGRAPH Comput. Graph., 41

\bibitem[{{Durech} {et~al.}(2010){Durech}, {Sidorin}, \&
  {Kaasalainen}}]{2010A&A...513A..46D}
{Durech}, J., {Sidorin}, V., \& {Kaasalainen}, M. 2010, \aap, 513, A46

\bibitem[{Elvidge {et~al.}(1997)Elvidge, Baugh, Kihn, Kroehl, \&
  Davis}]{Elvidge1997}
Elvidge, C.~D., Baugh, K., E., Kihn, E.~A., Kroehl, H.~W., \& Davis, E.~R.
  1997, Photogramm. Eng. Remote Sensing, 63, 727

\bibitem[{Elvins(1992)}]{Elvins:1992:SAV:142413.142427}
Elvins, T.~T. 1992, SIGGRAPH Comput. Graph., 26, 194

\bibitem[{Farouki(2012)}]{Farouki:2012:BPB:2221977.2222160}
Farouki, R.~T. 2012, Comput. Aided Geom. Des., 29, 379

\bibitem[{{Feng} {et~al.}(2011){Feng}, {Croft}, {Di Matteo}, {Khandai},
  {Sargent}, {Nourbakhsh}, {Dille}, {Bartley}, {Springel}, {Jana}, \&
  {Gardner}}]{2011ApJS..197...18F}
{Feng}, Y., {Croft}, R.~A.~C., {Di Matteo}, T., {et~al.} 2011, \apjs, 197, 18

\bibitem[{Frenkel(1988)}]{Frenkel:1988:ASV:42372.42373}
Frenkel, K.~A. 1988, Commun. ACM, 31, 111

\bibitem[{Friendly(2006)}]{Friendly:06:hbook}
Friendly, M. 2006, in Handbook of Computational Statistics: Data Visualization, Vol. III
  ed. C.~Chen, W.~H\"ardle, \& A.~Unwin (Heidelberg: Springer-Verlag), 1

\bibitem[{Garland {et~al.}(2008)Garland, Le~Grand, Nickolls, Anderson,
  Hardwick, Morton, Phillips, Zhang, \& Volkov}]{4626815}
Garland, M., Le~Grand, S., Nickolls, J., {et~al.} 2008, IEEE Micro, 28, 13

\bibitem[{Gooch(1995)}]{485155}
Gooch, R., 1995, in Proc. sixth Conference on Visualization (Washington, DC: IEEE Computer Society), 374 

\bibitem[{{Goodman}(2012)}]{2012AN....333..505G}
{Goodman}, A.~A. 2012, Astron. Nachr., 333, 505

\bibitem[{{Hansen} \& {Johnson}(2005)}]{Hansen2005}
{Hansen}, C.~D., \& {Johnson}, C.~R. eds. 2005, The Visualization Handbook (San Diedo: Academic Press)

\bibitem[{{Hassan} \& {Fluke}(2011)}]{2011PASA...28..150H}
{Hassan}, A., \& {Fluke}, C.~J. 2011, Proc. Astron. Soc. Australia, 28, 150

\bibitem[Hassan et al.(2012)]{2012PASA...29..340H} Hassan, A.~H., Fluke, 
C.~J., \& Barnes, D.~G.\ 2012, Proc. Astron. Soc. Australia, 29, 340 

\bibitem[{Hess(2010)}]{Hess:2010:BFE:1893021}
Hess, R. 2010, Blender Foundations: The Essential Guide to Learning Blender 2.6. (Burlington: Focal Press)

\bibitem[{Hunter(2007)}]{4160265}
Hunter, J. 2007, Comput. Sci. Eng., 9, 90

\bibitem[{Jacob \& Plesea(2001)}]{931430}
Jacob, J.~C., \& Plesea, L. 2001, in Proc. IEEE Aerospace Conf., Vol.~7, 7--3530

\bibitem[{{Kent}(2011)}]{2011ASPC..442..625K}
{Kent}, B.~R. 2011, in ASP Conf. Ser. 442, Astronomical Data Analysis Software and Systems XX, ed. I.~N.
  {Evans}, A.~{Accomazzi}, D.~J. {Mink}, \& A.~H. {Rots} (San Francisco: ASP), 625

\bibitem[{Kuchelmeister {et~al.}(2012)Kuchelmeister, M\"{u}ller, Ament, Wunner,
  \& Weiskopf}]{3c40fd5dd5}
Kuchelmeister, D., M\"{u}ller, T., Ament, M., Wunner, G., \& Weiskopf, D. 2012,
  Comput. Phys. Commun., 183, 2282

\bibitem[{Laramee(2008)}]{Laramee:2008:CEC:1364835.1364838}
Laramee, R.~S. 2008, Softw. Pract. Exp., 38, 735

\bibitem[{{Leech} \& {Jenness}(2005)}]{2005ASPC..347..143L}
{Leech}, J., \& {Jenness}, T.~J. 2005, in ASP Conf. Ser. 347, Astronomical Data Analysis Software and Systems
  XIV, ed. P.~{Shopbell}, M.~{Britton}, \& R.~{Ebert} (San Francisco: ASP), 143

\bibitem[{{Levay}(2011)}]{2011ASPC..442..169L}
{Levay}, Z. 2011, in ASP Conf. Ser. 442, Astronomical Data Analysis Software and Systems XX, ed. I.~N.
  {Evans}, A.~{Accomazzi}, D.~J. {Mink}, \& A.~H. {Rots} (San Francisco: ASP), 169

\bibitem[{Levoy(1988)}]{Levoy:1988:DSV:44650.44652}
Levoy, M. 1988, IEEE Comput. Graph. Appl., 8, 29

\bibitem[{Lip\c{s}a {et~al.}(2012)Lip\c{s}a, Laramee, Cox, Roberts, Walker,
  Borkin, \& Pfister}]{CGF:CGF3184}
Lip\c{s}a, D.~R., Laramee, R.~S., Cox, S.~J., Roberts, J.~C., Walker, R., Borkin, M.~A., \& Pfister, H. 2012, Comput. Graph.  Forum, 31, 2317

\bibitem[{Lorensen \& Cline(1987)}]{Lorensen:1987:MCH:37402.37422}
Lorensen, W.~E., \& Cline, H.~E. 1987, SIGGRAPH Comput. Graph., 21, 163

\bibitem[{{Lupton} {et~al.}(2004){Lupton}, {Blanton}, {Fekete}, {Hogg},
  {O'Mullane}, {Szalay}, \& {Wherry}}]{2004PASP..116..133L}
{Lupton}, R., {Blanton}, M.~R., {Fekete}, G., Hogg, D.~W., O'Mullane, W., Szalay, A., \& Wherry, N. 2004, \pasp, 116, 133

\bibitem[{{Mercer} \& {Klimenko}(2008)}]{2008CQGra..25r4025M}
{Mercer}, R.~A., \& {Klimenko}, S. 2008, Class. Quantum Gravity, 25, 184025

\bibitem[{Molnar {et~al.}(1994)Molnar, Cox, Ellsworth, \& Fuchs}]{291528}
Molnar, S., Cox, M., Ellsworth, D., \& Fuchs, H. 1994, Comput. Graph. Appl., 14, 23

\bibitem[{Moloney {et~al.}(2011)Moloney, Ament, Weiskopf, \&
  M\"{o}ller}]{tvcg11_dvr}
Moloney, B., Ament, M., Weiskopf, D., \& M\"{o}ller, T. 2011, IEEE Trans. Visual. Comput. Graph., 17, 1164

\bibitem[{Munzner {et~al.}(2006)Munzner, Johnson, Moorhead, Pfister, Rheingans \& Yoo}]{Munzner2006}
Munzner, T., Johnson, C., Moorhead, R., Pfister, H., Rheingans, P.,  \& Yoo, T.~S. 2006, IEEE Comput. Graph. Appl., 26, 20

\bibitem[{Murray \& vanRyper(1996)}]{0083492}
Murray, J.~D., \& van Ryper, W. 1996, Encyclopedia of Graphics File Formats (second ed.) (Paris: O'Reilly \& Associates, Inc.)

\bibitem[{Norman(2012)}]{Norman2012}
Norman, C. 2012, Science, 335, 525

\bibitem[{Patoli {et~al.}(2009)Patoli, Gkion, Al-Barakati, Zhang, Newbury, \&
  White}]{4839978}
Patoli, M., Gkion, M., Al-Barakati, A., Zhang, W., Newbury, P., \& White, M. 2009, in IEEE Power Systems
  Conference and Exposition (Seattle: IEEE), 1

\bibitem[{{Peng} {et~al.}(2010){Peng}, {Ruan}, {Long}, {Simpson}, \&
  {Myers}}]{Peng2010}
{Peng}, H., {Ruan}, Z., {Long}, F., {Simpson}, J.~H., \& {Myers}, E.~W. 2010,
  Nat. Biotechnol., 28, 348

\bibitem[{{Phong}(1975)}]{Phong:1975:ICG}
{Phong}, B. 1975, Commun. ACM, 18, 311

\bibitem[{Porter \& Duff(1984)}]{Porter:1984:CDI:964965.808606}
Porter, T., \& Duff, T. 1984, SIGGRAPH Comput. Graph., 18, 253

\bibitem[{Price {et~al.}(2011)Price, Puchala, Rovito, \& Priddy}]{6183117}
Price, R., Puchala, J., Rovito, T., \& Priddy, K. 2011, in Proc. IEEE National Aerospace and Electronics Conf. (Fairborn: IEEE), 291

\bibitem[Rana \& Herrmann(2011)]{2011PhFl...23i1109R} Rana, S., \& Herrmann, M.\ 2011, Phys. Fluids, 23, 091109 

\bibitem[{{Rector} {et~al.}(2007){Rector}, {Levay}, {Frattare}, {English}, \&
  {Pu'uohau-Pummill}}]{2007AJ....133..598R}
{Rector}, T.~A., {Levay}, Z.~G., {Frattare}, L.~M., {English}, J., \&
  {Pu'uohau-Pummill}, K. 2007, \aj, 133, 598

\bibitem[{Reniers \& Telea(2008)}]{Reniers2008}
Reniers, D. \& Telea, A., 2008, in IEEE International Conf. on Shape Modeling and Applications (Stony Brook: IEEE), 273

\bibitem[{Rhyne(2012)}]{Rhyne:2012:ACT:2343483.2343484}
Rhyne, T.-M. 2012, in ACM SIGGRAPH 2012 Courses (New York: ACM), 1:1--1:82

\bibitem[{Robinson(1982)}]{nla.cat-vn108090}
Robinson, A.~H. 1982, Early Thematic Mapping in the History of Cartography (Chicago: University of Chicago Press)

\bibitem[{Scianna(2013)}]{scianna2013}
Scianna, A. 2013, Appl. Geomatics, 1

\bibitem[{Shahidi {et~al.}(1996)Shahidi, Lorensen, Kikinis, Flynn, Kaufman, \&
  Napel}]{568151}
Shahidi, R., Lorensen, B., Kikinis, R., Flynn, J., Kaufman, A., \& Napel, S. 1996, in Proc. seventh IEEE Conf. on Visualization (San Francisco: IEEE), 439

\bibitem[{{Springel}(2005)}]{2005MNRAS.364.1105S}
{Springel}, V. 2005, \mnras, 364, 1105

\bibitem[{Staples \& Bieman(1999)}]{journals/ac/StaplesB99}
Staples, M.~L., \& Bieman, J.~M. 1999, Adv. Comput., 49, 95

\bibitem[{Stone {et~al.}(2010)Stone, Gohara, \& Shi}]{5457293}
Stone, J., Gohara, D., \& Shi, G. 2010, Comput. Sci. Eng., 12, 66

\bibitem[{Strand \& Borgefors(2005)}]{Strand2005}
Strand, R., \& Borgefors, G. 2005, Comput. Vision Image Understand., 100, 3

\bibitem[{Strehl \& Ghosh(2002)}]{Strehl02relationship-basedclustering}
Strehl, A., \& Ghosh, J. 2002, INFORMS J. Comput., 15, 2003

\bibitem[{SubbaRao {et~al.}(2008)SubbaRao, Arag\'{o}n-Calvo, Chen, Quashnock,
  Szalay, \& York}]{1367-2630-10-12-125015}
SubbaRao, M.~U., Arag\'{o}n-Calvo, M.~A., Chen, H.~W., Quashnock, J.~M., Szalay, A.~S., \& York, D.~G. 2008, New
  J. of Phys., 10, 125015

\bibitem[{{Taylor} {et~al.}(2011){Taylor}, {Davies}, \&
  {Minchin}}]{2011AAS...21840822T}
{Taylor}, R., {Davies}, J.~I., \& {Minchin}, R.~F. 2011, in AAS Meeting Abstracts 218, 408.22

\bibitem[{{Teuben} {et~al.}(2001){Teuben}, {Hut}, {Levy}, {Makino}, {McMillan},
  {Portegies Zwart}, {Shara}, \& {Emmart}}]{2001ASPC..238..499T}
{Teuben}, P.~J., {Hut}, P., {Levy}, S., Makino, J., McMillan, S., Portegies Zwart, S., Shara, M., \& Emmart, C. 2001, in ASP Conf. Ser. 238, Astronomical Data Analysis
  Software and Systems X, ed. F.~R. {Harnden}, Jr., F.~A. {Primini}, \& H.~E.
  {Payne} (San Francisco: ASP), 499

\bibitem[{Teyseyre \& Campo(2009)}]{Teyseyre:2009:OSV:1477065.1477374}
Teyseyre, A.~R., \& Campo, M.~R. 2009, IEEE Trans. Visual. Comput. Graph., 15, 87

\bibitem[{{Tully} {et~al.}(2009){Tully}, {Rizzi}, {Shaya}, {Courtois},
  {Makarov}, \& {Jacobs}}]{2009AJ....138..323T}
{Tully}, R.~B., {Rizzi}, L., {Shaya}, E.~J., Courtois, H.~M., Makarov, D.~I., \& Jacobs, B.~A. 2009, \aj, 138, 323

\bibitem[{{Turk} {et~al.}(2011){Turk}, {Smith}, {Oishi}, {Skory}, {Skillman},
  {Abel}, \& {Norman}}]{2011ApJS..192....9T}
{Turk}, M.~J., {Smith}, B.~D., {Oishi}, J.~S., Skory, S., Skillman, S.~W., Abel, T., \& Norman, M.~L. 2011, \apjs, 192, 9

\bibitem[Valverde \& Sol{\'e}(2005)]{2005PhRvE..72b6107V} Valverde, S., \& Sol{\'e}, R.~V.\ 2005, Phys. Rev. E, 72, 026107 

\bibitem[{Wainer \& Velleman(2001)}]{doi:10.1146/annurev.psych.52.1.305}
Wainer, H., \& Velleman, P.~F. 2001, Ann. Rev. Psychol., 52, 305

\bibitem[{Wallace(1991)}]{Wallace:1991:JSP:103085.103089}
Wallace, G.~K. 1991, Commun. ACM, 34, 30

\bibitem[{{Walter} {et~al.}(2008){Walter}, {Brinks}, {de Blok}, {Bigiel},
  {Kennicutt}, {Thornley}, \& {Leroy}}]{2008AJ....136.2563W}
{Walter}, F., {Brinks}, E., {de Blok}, W.~J.~G., Bigiel, F., Kennicutt, R.~C., Jr., Thornley, M.~D., \& Leroy, A. 2008, \aj, 136, 2563

\bibitem[{Wenger {et~al.}(2012)Wenger, Ament, Guthe, Lorenz, Tillmann,
  Weiskopf, \& Magnor}]{nebulae-vis2012}
Wenger, S., Ament, M., Guthe, S., Lorenz, D., Tillmamm, A., Weiskopf, D., \& Magnor, M. 2012, IEEE Trans.
  Visual. Comput. Graph., 18, 2188

\bibitem[{Yagel(1996)}]{Yagel96classificationand}
Yagel, R. 1996, SIGGRAPH Tutorial Notes, Course No. 34 (New Orleans: ACM)

\bibitem[{Yost \& North(2006)}]{4015437}
Yost, B., \& North, C. 2006, IEEE Trans. Visual. Comput. Graph., 12, 837

\bibitem[{{Zini} {et~al.}(2010){Zini}, {Porozov}, {Andrei}, {Loni}, {Caudai},
  \& {Zopp{\`e}}}]{2010arXiv1009.4801Z}
{Zini}, M.~F., {Porozov}, Y., {Andrei}, R.~M., Loni, T., Caudai, C., \& Zopp\'{e}, M. 2010, preprint (arXiv:1009.4801)

\end{thebibliography}

\clearpage


\begin{deluxetable}{lccc}
\tablecolumns{4}
\tablecaption{Blender Data Formats\label{blenderformats}}
\tablewidth{0pc}
\tabletypesize{\scriptsize}
\tablehead{ 
	\colhead{\textbf{Format}} & 
	\colhead{\textbf{Type}} &
	\colhead{\textbf{Suffix}} & 
	\colhead{\textbf{Usage}}

}
\startdata
Blend &  native                & .blend       & Import/Export            \\
OBJ   &  ASCII                 & .obj         & Import/Export            \\
CSV   &  ASCII                 & .csv         & Import\tablenotemark{a}  \\
JPEG  &  image                 & .jpg, .jpeg  & Import\tablenotemark{b} / Export    \\
TIF   &  image                 & .tif, .tiff  & Import\tablenotemark{b} / Export    \\
PNG   &  image                 & .png         & Import\tablenotemark{b} / Export    \\
GIF   &  image                 & .gif         & Import\tablenotemark{b} / Export    \\
EXR   &  image                 & .exr         & Import                   \\
FITS  &  image/structured data & .fit, .fits  & Import\tablenotemark{a}  \\
AVI   &  video                 & .avi         & Export                   \\
MPEG  &  video                 & .mpg, .mpeg  & Export                   \\
MOV   &  video                 & .mov         & Export\tablenotemark{c}  \\
H.264 &  video                 & .mov         & Export\tablenotemark{c}  \\
\enddata
\tablecomments{We provide examples and scripts in Section \ref{summary}.
}
\tablenotetext{a}{Can be imported via standard Python file I/O.}
\tablenotetext{b}{Can be used for surface mesh texturing.}
\tablenotetext{c}{Can be achieved on Linux-based OS via FFmpeg (http://www.ffmpeg.org)}
\end{deluxetable}

\begin{deluxetable}{lcccc}
\tablecolumns{5}
\tablecaption{Rendering Time Benchmark Comparison\label{gputable}}
\tablewidth{0pc}
\tablehead{ 
	\colhead{\textbf{Render}} & 
	\colhead{\textbf{Samples}} &
	\colhead{\textbf{Mean}} &
	\colhead{\textbf{Median}} & 
	\colhead{\textbf{SE$_{\bar{x}}$}}
	
	\\
	
	\colhead{} &
	\colhead{\textit{N}} &
	\colhead{minutes} &
	\colhead{minutes} &
	\colhead{ } 
}
\startdata
NVidia CUDA &	100 &	2.82 &	1.49 & 0.31 \\
OpenCL &	31 &	2.17 &	1.99 & 0.22 \\
CPU &		154 &	8.30 &	6.48 & 0.47 \\
\enddata
\tablecomments{Rendering time for a Blender session in the Cycles engine at a resolution of 960 $\times$ 540 pixels with rendering tile sizes of 8 $\times$ 8 pixels.   The benchmark test consisted of 6 camera views and 12 mesh objects totaling 86,391 polygons and 91,452 vertices.  The last column describes the standard error of the mean.
}
\end{deluxetable}

\begin{deluxetable}{lccccc}
\tablecolumns{6}
\tablecaption{Comparison of 3D Graphics Software and Engines\label{softwarecompare}}
\tablewidth{0pc}
\tabletypesize{\scriptsize}
\tablehead{ 
	\colhead{\textbf{Format}} & 
	\colhead{\textbf{Platform}\tablenotemark{a}} &
	\colhead{\textbf{License/Avail}} & 
	\colhead{\textbf{Features}\tablenotemark{b}} &
	\colhead{\textbf{Free?}\tablenotemark{c}} &
	\colhead{\textbf{Reference}}

}
\startdata
Blender         &  L,M,W        &  GPL          & MM, A, T, iR, eR, L, S, VE, N & Yes & 1       \\
3D Studio Max   &  W            &  Proprietary  & MM, A, T, iR, eR(some), L     & No  & 2       \\
Cinema 4D       &  L,M,W        &  Proprietary  & MM, A, T, L, eR               & No  & 3       \\
EIAS3D          &  M,W          &  Proprietary  & MM, A, T, L, iR               & No  & 4       \\
Houdini         &  L,M,W        &  Proprietary  & MM, A, T, L, eR               & No  & 5       \\
Lightwave 3D    &  M,W          &  Proprietary  & MM, A, T, L, iR               & No  & 6       \\
Maya            &  L,M,W        &  Proprietary  & MM, A, T, L, iR,              & No  & 7       \\
Modo            &  M,W          &  Proprietary  & MM, A, T, L                   & No  & 8       \\
SAP V.E.S.      &  W            &  Propreitary  & MM, A, iR                     & No  & 9      \\
SoftImage       &  L,W          &  Propreitary  & MM, A, S, N                   & No  & 10      \\

\enddata
\tablecomments{References and Vendor websites: \\
(1) http://www.blender.org/ \\
(2) http://www.autodesk.com/products/autodesk-3ds-max/overview/ \\
(3) http://www.maxon.net/ \\
(4) http://www.eias3d.com/ \\
(5) http://www.sidefx.com/ \\
(6) https://www.lightwave3d.com/ \\
(7) http://www.autodesk.com/products/autodesk-maya/overview/ \\
(8) http://www.luxology.com/modo/ \\
(9) Visual Enterprise Solutions \\ including Client View and Deep Server: http://www.sap.com/ \\
(10) http://www.autodesk.com/products/autodesk-softimage/overview/ \\
}
\tablenotetext{a}{M: Mac OS X, L: Linux, W: Windows}
\tablenotetext{b}{MM: 3D Mesh models, A: animation control, T: 2D/3D texture mapping, iR: internal rendering, eR: support for external rendering, L: lighting control, S: 3D sculpting, VE: video editing, N: Node editing}
\tablenotetext{c}{Some commercial software vendors offer free trials.}

\end{deluxetable}

\clearpage

\end{document}